\documentstyle[prb,aps,epsfig,multicol]{revtex} 
 
\newcommand{\nvec}{\mbox{$\underline{n}$}} 
\newcommand{\kvec}{\mbox{$\underline{k}$}} 
\newcommand{\rvec}{\mbox{$\underline{r}$}} 
\newcommand{\vvec}{\mbox{$\underline{v}$}} 
\newcommand{\uvec}{\mbox{$\underline{u}$}} 
\newcommand{\ucoarse}{\mbox{$\underline{\cal U}$}} 
\newcommand{\Mmat}{\mbox{$\underline{\underline{M}}$}} 
 
\begin{document} 
\title{Continuum limit of amorphous elastic bodies:\\ 
A finite-size study of low frequency harmonic vibrations} 
%Non-affine {\em vs.} affine displacement fields} 
\author{ 
A.~Tanguy\thanks{E-mail: atanguy@dpm.univ-lyon1.fr}, J.P. Wittmer, 
F.~Leonforte \and J.-L.~Barrat} 
 
\address{ 
D\'epartement de Physique des Mat\'eriaux, 
Universit\'e Claude Bernard \& CNRS\\ 
69622 Villeurbanne Cedex, France} 
 
\date{\today} 
\setcounter{page}{1} 
\maketitle 
 
\begin{abstract} 
The approach of the elastic continuum limit in small amorphous 
bodies formed by weakly polydisperse Lennard-Jones beads is 
investigated in a systematic finite-size study. We show that 
classical continuum elasticity breaks down when the wavelength of 
the sollicitation is smaller than a characteristic length of 
approximately 30 molecular sizes. Due to this surprisingly large 
effect ensembles containing up to $N=40,000$ particles have been 
required in two dimensions to yield a convincing match with the 
classical continuum predictions for the eigenfrequency spectrum of 
disk-shaped aggregates and periodic  bulk systems. The existence 
of an effective length scale $\xi$ is confirmed by the analysis of 
the (non-gaussian) noisy part of the low frequency vibrational 
eigenmodes. Moreover, we relate it  to the {\em non-affine} part 
of the displacement fields under imposed elongation and shear. 
Similar correlations (vortices) are indeed observed on distances 
up to $\xi\approx 30$ particle sizes. 

\end{abstract} 
\centerline{Pacs: 72.80.Ng, 65.60.+a, 61.46.+w} 
 
\begin{multicols}{2} 
 
\newpage 
\section{Introduction.}\label{sec:Intro} 
 
%\fbox{Key paragraph: Limits of continuum theory} 
 
Determining the vibration frequencies and the associated 
displacement fields of solid bodies with various shapes is a well 
studied area of continuum mechanics 
\cite{elastodynamics,landau,lubensky} with applications in fields 
as different as planetary science and nuclear physics. 
The increasing development of materials containing nanometric size structures 
leads one to question the limits of applicability of classical continuum elasticity theory, 
which is in principle valid only on length scales much larger than the 
interatomic distances \cite{lubensky,alexander}. 
This question is relevant from an experimental viewpoint, since 
mechanical properties are inferred from spectroscopic measurements 
systematically interpreted within the framework of continuum 
elasticity \cite{vallee,nanopart,hodak,saviot}. As increasingly 
smaller length scales are now investigated \cite{saviot}, direct 
verification of this assumption is highly warranted. 
 
For macroscopic systems, on the other side, it is well known that 
the vibrational density of states in amorphous glassy materials 
deviates from the classical spectrum at the so called '' Boson 
peak`` frequency which is in the Terahertz 
range~\cite{boson,glassyvibs1,Schirmacher,Ruocco}. The nature of 
the ''Boson peak`` is highly 
controversial~\cite{localization,ruocco2}. However, this 
experimental fact suggests that continuum theory is inappropriate 
at small length scales where the disorder of the amorphous system 
may become relevant~\cite{EPLpaper}. Obviously, one may ask if 
there is a finite length scale below which the classical 
mechanical approach becomes inappropriate, and what the 
microscopic features are which determine 
it~\cite{alexander,localization,ruocco2,diffusion}. Elaborating 
further the brief presentation given in~\cite{EPLpaper}, we show 
by means of a simple generic simulation model that, indeed, a 
relatively large characteristic length exists, and, second, that 
it envolves collective particle rearrangements. 
 
The above questions are more generally  related to the propagation 
of waves in disordered materials~\cite{sheng}, and concerns 
foam~\cite{mousses} and emulsions~\cite{langer} as well as 
granular materials~\cite{leibig,melin,radjai}, when they are 
submitted to small amplitude vibrations. As for these systems of 
current interest, the existence of an elastic limit is still a 
matter of debate \cite{chaos}, we believe that the detailed 
characterization of strongly heterogeneous elastic benchmark 
systems with a well defined continuum limit approach is crucial. 
 
%\fbox{What we did} 
 
In this paper, we investigate the existence of a continuum  limit 
in the vibrational modes of two-dimensional amorphous nanometric 
Lennard-Jones materials. The objects we consider are either 
disk-shaped clusters of diameter $2R$, as the one shown on the 
left side of Fig.~\ref{figFlink}, or bulk like systems without 
surfaces contained in a square of side $L$ with periodic boundary 
conditions (Fig.~\ref{figFlink}(b)). Technically, the systems are 
formed by carefully quenching a slightly polydisperse liquid of 
spherical particles interacting via simple Lennard-Jones (LJ) pair 
potentials into the nearest energy minimum. 
Due to the polydispersity the resulting structures are isotropic 
and {\em amorphous}, i.e. exhibit no long range crystalline order. 
The force network (Fig.~\ref{figFlink}) appears to be strongly 
varying with weak and tensile zones (red) embedded within a rigid 
repulsive skeleton (black). These ''force chains`` are very 
similar to those found in cohesionless granular media without 
attractive forces \cite{forcechains}. This feature may be added to 
the list of similarities which have been noticed between granular 
and amorphous (glassy) materials \cite{chaos,liunagel,langer}. As 
the force network is strongly inhomogeneous, the relevance of the 
quenched stresses is a natural question~\cite{alexander}. 
 
We investigate the vibrational modes of these objects using atomic level simulations. 
All particle coordinates and interparticle forces are exactly known here, and 
it is possible to calculate the vibration frequencies around an equilibrium 
position, by exact diagonalization of the so called dynamical 
matrix \cite{Kittel} expressible in terms of the first and second derivatives 
of the interparticle interaction potentials. 
We have carried out a systematic comparison of these eigenfrequencies $\omega(p)$ 
($p$ being an index increasing with frequency) obtained numerically 
with those predicted by continuum elasticity for two 
%(and three) 
dimensional objects of increasingly large 
sizes~\cite{elastodynamics}. We concentrate on the lowest end of 
the vibrational spectrum, since this is the part that corresponds 
to the largest wavelengths for the vibrations,  and should reach 
first the continuum limit. These frequencies are also those which 
are probed in low frequency Raman scattering experiments 
\cite{saviot}, in order to determine the typical size of 
nanoparticles. 
 
%\fbox{Key results} 
 
The key result of this paper is to show  the existence of a 
characteristic wavelength (thus a characteristic size) beyond 
which the classical continuum limit is valid, but below which it 
is erroneous. Moreover, we show the existence of rotational 
structures (vortices) of similar sizes, when the system is 
submitted to simple mechanical sollicitations (traction and 
shear). The size of these vortices is relatively large ($\approx 
30$ average interatomic distances). We discuss the relation 
between the sizes of the vortices and the limit of applicability 
of the classical continuum theory by computing the elastic moduli, 
by studying the symmetry of the nanoscale stress tensor, and by 
identifying the low frequency vibrational eigenmodes.

%\fbox{Outline} 
 
Our paper is arranged as follows: In Sec.~\ref{sec:Theory} we 
summarize  some basic relations and results of classical continuum 
theory for two dimensional elastic bodies. Simulation techniques, 
sample parameters and preparation protocols of our model amorphous 
systems are explained in Sec.~\ref{sec:Samples} and simple system 
properties are discussed. 
In  Sec.~\ref{sec:Quench} we analyse histograms and spatial 
correlations of the quenched forces and of the stiffness of the 
bonds. 
A weak enhancement of the rigid skeleton is demonstrated for small systems. 
The next two sections contain the key results of this paper. 
In Sec.~\ref{sec:Nonaffine} we discuss the mechanical properties 
of a periodic bulk system under elongation and shear, we compute 
the elastic moduli and we characterize the non-affine displacement 
field generated. 
The eigenvalues and eigenvectors and their departures from the continuum prediction 
are analyzed in Sec.~\ref{sec:Eigenmodes}. 
We conclude with a summary of our results in Sec.~\ref{sec:Discussion}. 

\section{Theory}\label{sec:Theory}

Before presenting our computational results we give here a short
synopsis of assumptions and results of classical continuum
elasticity in two dimensions, relevant for the subsequent
discussion of our numerical results.

\subsection{Continuum description of an isotropic elastic body in 2D.}\label{sec:continuum}

The main assumption of classical elasticity~\cite{salencon}, is
that the system can be entirely described by a unique vectorial
field: the displacement field $\uvec(\rvec)$, describing the
displacement of a volume element from its equilibrium position
$\rvec$. As we are interested in systems at zero temperature,
energy and free energy are obviously identical. The Landau
expansion\cite{landau} of the energy $\delta E$ per unit volume is
expressed in terms of $\uvec$ and its first derivatives. Due to
translational and rotational invariance~\cite{salencon} of $\delta
E$, it depends only up to second order on the symmetric part of
$\mathrm{\mathbf{grad }}\,\, \uvec$, the (linearized) strain
tensor $\epsilon_{\alpha\beta}\equiv 1/2 ~ (\partial
u_{\alpha}/\partial x_{\beta} +
\partial u_{\beta}/\partial x_{\alpha})$ with $x_1=x$ and $x_2=y$ for the coordinates.
Moreover in linear elasticity (that is in the lowest order in
the Landau expansion), for isotropic and homogeneous systems only
two Landau parameters are required, thus:
\begin{equation}
\delta E = \frac{\lambda}{2} \left( \epsilon_{xx}+\epsilon_{yy} \right)^2
         + \mu \left( \epsilon_{xx}^2+\epsilon_{yy}^2+2\epsilon_{xy}^2 \right)
\label{eq:Landau}
\end{equation}
where we have defined the two phenomenological Lamé coefficients
 $\lambda$ and $\mu$. $\mu$ is the so called shear modulus,
and $\lambda$ takes part in the compressibility $k = \lambda +
\mu$. Below, we will discuss two methods to determine them in a
computer experiment. The stress tensor can be defined as the
conjugate variable of the strain tensor,
\begin{equation}
\label{eq:stress}
\sigma_{\alpha\beta}\equiv \partial \delta E/ \partial \epsilon_{\alpha\beta}.
\end{equation}
In this definition $\sigma_{\alpha\beta}$ it obviously  a
symmetric tensor. We will see later that a microscopically
constructed stress tensor can, however, violate this symmetry
condition at small length scales (section ~\ref{sec:stresspart}).

For the remaining three independent stress tensor elements, one
gets from Eqns~(\ref{eq:Landau}) and (~\ref{eq:stress}) the
standard Hooke 's relations:
\begin{eqnarray}
\label{eq:Hooke}
\sigma_{xx} & = & (\lambda + 2 \mu) \epsilon_{xx} + \lambda \epsilon_{yy}  \\ \nonumber
\sigma_{yy} & = & (\lambda + 2 \mu) \epsilon_{yy} + \lambda \epsilon_{xx} \\ \nonumber
\sigma_{xy} & = & \sigma_{yx} = 2 \mu \epsilon_{xy}.
\end{eqnarray}
We use these formulae in Sec.~\ref{sec:Nonaffine} to measure directly the
Lam\'e coefficients from the forces generated in a periodic box under external strain.
It is easy to work out from Eq.~(~\ref{eq:Hooke}) that in 2
dimensions the Poisson ratio is given by
$\nu=\lambda/(\lambda+2\mu)$ with an upper bound $1$ (and not
$1/2$ as in 3D) \cite{landau}.
%

%\fbox{Vibrations.}

As it is well known\cite{elastodynamics,landau}, the equations of
motion $\rho \partial \sigma_{\alpha \beta} / \partial x_{\alpha}
= \ddot{u}_{\alpha}$ together with the constitute
Eqns.~(~\ref{eq:Hooke}) correspond to wave equations with boundary
conditions depending on the problem of interest: for the periodic
box problem the solutions must have the same periodicity, and the
freely floating disk-shaped aggregate requires vanishing lateral
and radial stresses on its surface.
The velocities of transverse and longitudinal waves
which solve these wave equations are given in terms of the Lam\'e coefficients,
i.e. one has $c_T^2=\mu/\rho$ and
$c_L^2=(\lambda+2\mu)/\rho$ where $\rho$ is the particle density \cite{elastodynamics}.
We remind that $c_L > c_T$ and that transverse modes
thus correspond generally to smaller eigenfrequencies.

The solutions for the periodic box case are of course the plane
waves with a wavevector quantified by the boundary conditions, $(k_x,k_y) = \frac{2\pi}{L} (n,m)$,
with wavelength~\cite{Note}
\begin{equation}
\lambda(p)=\lambda(n,m)=\frac{2\pi}{||\kvec||}=\frac{L}{(n^2+m^2)^{1/2}}
\label{eq:lambda}
\end{equation}
and with dimensionless frequency
\begin{equation}
\Omega^2_{T,L}(p) \equiv \left( \frac{\omega(p) L}{2\pi c_{T,L}} \right)^2 = n^2 + m^2
\label{eq:evperiodic}
\end{equation}
with two quantum numbers $n,m=0,1,...$. The running index $p$
increases with frequency.  In the continuous case, the dispersion
relation is linear, and the frequency is straightforward. Hence,
eigenfrequencies are characterized by a pair of different
integers. They are eight-fold degenerated if $n \ne m \ne 0$ and
four-fold in all other cases.
%(Note that so-called `accidental' higher degeneracies may occur.)
The associated plane waves travel in two opposite and orthogonal directions.
%

%\fbox{Eigenmodes for disks}

The situation for disk-shaped objects is somewhat more complex~\cite{elastodynamics},
with again two quantum numbers $n$ and $k$ characterizing the eigenmodes.
The quantum number $k$ is
associated with the angular dependency of the displacement field
(and is due to the $2\pi$ periodicity), and the
number $n$ to its radial dependency.
The eigenfrequencies are obtained by solving the non linear
dispersion relation. They are of the form
\begin{equation}
\Omega^2_{T,L}(p) \equiv \left( \frac{\omega(p) 2R}{2\pi c_{T,L}}\right)^2 = f_{nk}(\nu)
\nonumber
\end{equation}
where $\nu$ is the Poisson ratio given above.
The eigenvectors are related to Bessel functions~\cite{elastodynamics} which
may be approximated locally by plane waves if $\lambda(p) \ll R$.
Vibrational modes of disks are either non degenerate (for axially
symmetric modes) or  have two-fold degeneracy (for all other
modes). Indeed, for $k>0$, for every solution whose amplitude is
$\propto e^{ik\theta}$, one finds a  second  solution, orthogonal
to the first one,  with amplitude $\propto
e^{ik(\theta+\Delta\theta)}$ where $\Delta\theta=\pi/2k$. Every
additional solution with the same $k$ is a linear combination  of
these two vectors. For axially symmetric modes ($k=0$), the above
argument does not apply; any additional solution found by turning
the coordinate system is identical to the first one.

We finally stress the obvious: degenerate eigenvalues are inherent to the continuum
treatment of highly symmetric systems.
The failure to observe them indicates either the lifting of the relevant
symmetry or the breakdown of continuum theory.

\subsection{Pair potential systems with central forces.}

%\fbox{Dynamical matrix}

In a simulation one has the advantage to know all the individual
contributions to the total energy. The situation is particularly simple
if one has to deal with interparticle pair potentials $U(r_{ij})$
($r_{ij}$ being the interparticle distance) such as the LJ potential we use, and if we stay at zero temperature.
In this case, the difference of the total potential energy
\begin{equation}
\label{eq:potentiel}
E_p=\sum_{i=1}^{N-1}\sum_{j>i}^{N} U(r_{ij})
\end{equation}
due to a displacement field $\uvec$ can be written to second order as a Hessian form
$\delta E_p = \frac{1}{2} \uvec^{t} \cdot \Mmat \cdot \uvec$ in terms of the
$( 2N ) \times ( 2N )$ {\em dynamical matrix} $\Mmat$ whose elements are given by
\begin{equation}
 - r_{ij}^2 M_{i\alpha,j\beta} =
   r_{ij} t_{ij} \left( \delta_{\alpha\beta} - n_{\alpha} n_{\beta} \right)
 + r_{ij}^2 c_{ij} n_{\alpha} n_{\beta}
\label{eq:Dynamicmatrix}
\end{equation}
$\nvec$ being the unit vector of the bond (for simplicity, we do
not indicate the dependence of $\nvec$ on the particle indices $i$
and $j$), $t_{ij}\equiv\partial U(r_{ij})/\partial r_{ij}$ the
tension and $c_{ij}\equiv\partial^2U(r_{ij})/\partial r_{ij}^2$
the stiffness of the bond between two interacting beads $i$ and
$j$. The first one is related to the stresses ``quenched'' in the
bulk, but as we will see later it is in general small in
comparison with the second one.
As in the present study the mass $m$ of each monomer is set equal
to unity (even though the particle diameters are polydisperse) the
Euler-Lagrange equation $\Mmat \cdot \uvec + m \ddot{\uvec} = 0$
is solved directly by eigenfrequencies and displacement fields
obtained simply by diagonalisation of the dynamical matrix.

Assuming {\it constant} deformations  (purely affine displacement
field) under elongation and shear, one may use the dynamical
matrix to calculate the Lam\'e coefficients.
Comparing~\cite{hoover}  the Landau expression
Eq.~(~\ref{eq:Landau}) and the microscopic energy information in
Eq.~(~\ref{eq:potentiel}), we get
\begin{eqnarray}
\lambda_a & = &  \frac{1}{A} \sum_{ij}
\left[
r_{ij}   t_{ij}  n_x^2 n_y^2  +
r_{ij}^2 c_{ij}  \right( (n_x^4 + n_y^4)/2 - 2 n_x^2 n_y^2 \left)
\right]   \nonumber  \\
\mu_a     & = &  \frac{1}{A} \sum_{ij} r_{ij}^2 c_{ij} n_x^2 n_y^2
\label{eq:lambdamuaffine}
\end{eqnarray}
with $A$ being the total surface. The sums run over all pairs of
particles. As an {\it affine} displacement field is assumed to be
valid down to atomic distances the coefficients have been assigned
an index $a$ to distinguish them from the true macroscopic  Lam\'e
coefficients that will be calculated later.
It is instructive to consider the simple one dimensional analogy,
i.e. a long string of connected springs with force constants $c_l$,
to appreciate the approximation made in the above equations.
We recall that the effective macroscopic Young modulus is $\lambda = 1/\langle 1/c_l \rangle$
rather than  $\lambda_a = \langle c_l \rangle$. This is due to the fact that the
tension along the chain is constant but {\em not} the displacement field.
Hence, Eqns.~(~\ref{eq:lambdamuaffine}) are in general only good approximations
if the dispersion of the effective spring constants is small
and an external macroscopic strain is locally transmitted in an affine way.
($\lambda$ is then indeed given by the mean coupling constant with
a {\em negative} quadratic dispersion correction in leading order.)
We will test this assumption in Sec.~\ref{sec:Nonaffine} and
estimate the length below which the affinity approximation becomes
problematic.

\subsection{Definition and Measurement of stress tensor.}\label{sec:stresspart}

In the elastic continuum theory (section~\ref{sec:continuum}), the
stress tensor is usually introduced  by considering  the forces
per unit length exerted between adjacent volume elements. The
total energy $\delta E$ thus corresponds to the work of internal
forces. At a microscopic level, a natural way of extending this
macroscopic approach is to   compute $\sigma_{\alpha\beta}$  as
the sum of forces per unit length $(-t_{ij}n_\alpha /b)$ exerted
in the $\alpha$ direction through the side perpendicular to the
$\beta$ direction of a small  volume element of size $b$. A stress
tensor defined in this way can be asymmetric. The usual,
macroscopic  proof~\cite{landau,lubensky} of the symmetry of
$\sigma$ is based on the fact that  intermolecular forces are
short-ranged. Hence, one expects symmetry of the microscopic
stress tensor (obtained from the 'macroscopic' definition) only
for volume elements of size $b$ much larger than the range of
intermolecular forces. In section ~\ref{sec:Nonaffine}, we will
investigate how this macroscopic limit is reached.

Note that our definition of the stress tensor,  does {\it not}
correspond strictly to the usual~\cite{alexander,goldenberg}
microscopic Kirkwood definition
\begin{equation}
\nonumber
\sigma_{\alpha\beta}(i)\equiv -\Sigma_j t_{ij} n_\alpha.n_\beta
\end{equation}
which is necessarily symmetric. Although both quantities yield the
same macroscopic stress tensor, the definition we are using is
more appropriate to illustrate deviations from macroscopic
behavior at small scales.
Another method for illustrating these deviations can be found in
Ref.~\cite{goldenberg}. In this work, the local breakdown of
rotational invariance in a disordered system was shown to imply
contribution of the asymmetric part of $\uvec$ to the local
elastic energy. Defining $\sigma_{\alpha\beta}$ as $\partial\delta
E/\partial (\partial u_\alpha/\partial\beta)$ in place of
Eq.~(~\ref{eq:stress}), it can be shown that the contribution of
the asymmetric part of $\sigma$ to the local energy is related to
the work of the fluctuating part of the forces on the
fluctuating part of $\uvec$. In reference ~\cite{goldenberg}, it
was shown that  this contribution vanishes only for very large
systems.

\section{Sample preparation and characterization.}\label{sec:Samples}

%\fbox{Type of Systems and protocols.}

Systems with two different boundary conditions (disks and periodic bulk systems)
and three quench protocols have been simulated.
In this section we discuss some technical points concerning the simulation
methods and the sample preparation protocols and parameters.
The details of the protocols and some properties of the final configurations
are summarized in the Tables~\ref{tab:protI},~\ref{tab:protII} and ~\ref{tab:protIII}.

%\fbox{Potential}

In the present study we use a shifted LJ potential \cite{AT}
for polydisperse particles with
$U_{LJ}(r)=\epsilon \left( (r_0/r)^{12} - 2(r_0/r)^{6} \right) + const$ for
$r< 2r_0$ and $U_{LJ}(r)=0$ otherwise, correctly regularized.
We have written the potential in terms of the distance $r_0 \equiv (a_i+a_j)/2^{5/6}$
where the LJ force vanishes.
Natural LJ units are used, i.e. we set the energy parameter $\epsilon \equiv 1$,
the particle mass $m\equiv 1$ and the mean diameter $a=\langle a_i \rangle=1$.
Note that while the particle mass is strictly monodisperse
the particle diameters $a_i$ are homogeneously distributed between $0.8$ and $1.2$.
This corresponds to a polydispersity index $\delta a/a \approx 0.12$
which is sufficient to prevent large scale crystalline order.
We did not attempt to make the particles even more polydisperse fearing demixing
or systematic radial variation of particle sizes in the case of disk-shaped aggregates.

%\fbox{Simulation methods.}

The quench generally starts with molecular dynamics (MD) at some
fixed temperature using a simple velocity rescaling thermostat
\cite{AT}. As integrator we use the velocity Verlet algorithm
\cite{Thijssen} with a time increment $\delta t=0.001$. The
temperature remains constant over a fixed time interval of $1000$
MD steps for each temperature step.
This was sufficient to relax systems into a steady-state
(obviously, not necessarily the equilibrium)
as monitored by pressure and system energy.
Unfortunately, no more detailed characterization of the ageing as a function
of the quench rate has been recorded.
Following the initial MD sequence we quench the systems further
down using  a steepest descent (SD) method with a frictional force
proportional to the total particle velocity and an additional
damping term proportional to the relative velocity of interacting
particles. We use a small value of order one for both friction
constants and a simple leap-frog like integrator consistent with
the velocity depended force \cite{Thijssen}. The particle motion
is overdamped.
The systems were cooled down with SD over a period of at least $1000$ time steps.
Finally, the conjugate gradient method (CG) \cite{Thijssen,numrec}
was iterated till the configurations reach their local energy minima.
Double precision was used at every stage.
%

%\fbox{Disks.}

An example for a disk-shaped aggregate has already been presented
in fig.~\ref{figFlink}(a).
Two different protocols have been employed to generate such disks:
\begin{itemize}
\item
Protocol I: The starting point of the first protocol is a not too dense LJ liquid
droplet of radius $R_i$ and $\rho_i \approx 0.9$ and temperature $T_i=0.1$.
The initial radius $R_i$ is chosen such that the final mean radius
$R$ becomes not too different.
We cool the systems with MD steps at $T=0.1,0.05,0.01,0.005$ and $0.001$.
As described above, we finally quench each system into its local minimum
using a sequence of SD and CG steps. See Table~\ref{tab:protI} for details.
%We have in this way generated clusters with diameters up to $2R=220$
%and particle numbers between $N=16$ and $N=32,768$.
%
\item
Protocol II: In the second case spherical disks of radius $R_i$
were cut out of amorphous periodic systems already prepared at
$T=0$  (see Protocol III) and $\rho=0.925$  and the disks are
quenched as before with SD and CG. No finite temperature MD step
was included here. See Table~\ref{tab:protII} for details. The
second protocol is much faster than the first one which might,
however, mimic better the actually occurring aggregation process
in real systems.
\end{itemize}

The radius $R = \sqrt{2(I_1+I_2)/N}$ of quenched cluster
is obtained from
the eigenvalues $I_1 > I_2$ of the inertia tensor of the cluster.
The mean density $\rho = < N/\pi R^2>$ obtained accordingly
is a weakly decreasing function of the particle number as can be seen
from the Tables~\ref{tab:protI} and \ref{tab:protII}.
This is expected and in qualitative agreement with the
decreasing Laplace pressure $P\propto 1/R$.
Note that the pressure is very small for all the disks
(Protocol I and II alike) and is not indicated.
We have also tried to characterize the shape of the disks
and have indicated the excentricity $e=\sqrt{1-I_2/I_1}$.
As can be seen, small aggregates are strongly elliptic,
but the effect is much stronger for the first slow aggregation
protocol. This is probably due to capillary waves formed at
$T=0.1$  which are subsequently frozen in.
Obviously, ellipticity is one possible cause for lifting
of the eigenfrequency degeneracy. This effect should, however,
become small for larger clusters where $e$ vanishes
(Tab.~\ref{tab:protI} and \ref{tab:protII}).
Moreover, we will see later that  the mechanical properties of our
aggregates do not depend significantly on the quench protocol.

%\fbox{Periodic systems.}

Periodic bulk systems, such as the one presented in
Fig.~\ref{figFlink}(b), have been prepared following the third
protocol and are summarized in Table~\ref{tab:protIII}. We started
here with equilibrated liquids at $T=1$ which were then
subsequently cooled down with temperature steps at $T=0.5, 0.1,
0.05, 0.01, 0.005$ and $0.001$. Systems between $L=7a$ and
$L=208a$ and containing from $N=50$ up to $N=40,000$ particles
have been sampled.

For $N=10,000$ particles we have systematically scanned over density
varying the box size from $L=100$ to $L=110$.
This was done in order to find  --- for the given protocol ---
a working point density for which large bulk systems correspond to a near zero pressure
state $P(T=0)\approx 0$, thus the mechanical properties of the bulk systems
correspond to the free-floating aggregates.
Note that only one configuration has been sampled for this measurement sequence.
As a sideline, we draw attention to the fact that systems with $P < 0$
while mechanically stable in a periodic box are thermodynamically unstable
at low temperature.
If more time would be given to the systems to equilibrate as allowed by the protocol
phase separation would occur. Indeed we find evidence for the formation of
small holes for $\rho < 0.85$. In all other cases we find that density is perfectly
homogeneous down to a scale typical of the interatomic distance and density
fluctuations are easily excluded as a microscopic candidate for explaining
the slow continuum approach.
Our systematic finite-size study, i.e. the variation of box sizes $L$,
was performed at fixed density $\rho=0.925$.
In order to scale correctly the eigenfrequencies of the aggregates
we have also recorded the Lam\'e coefficients as a function of density.
We have checked that the particle energies $E$,
Lam\'e factors $\lambda_a$
and $\mu_a$, mean effective spring constants $\langle r_{ij}^2 c_{ij} \rangle$
(which are included in the different tables) do not depend significantly
on the different symmetries and quench protocols, for a given density.
It is reassuring that
all the results we obtained are robust with regard to the variation of the quench protocol
even though history dependence must obviously play a role for the details.
Typically, we have generated $20$ configurations for each $L$ and in terms of
CPU hours this was the most demanding part of this study. The simulations have
been performed on a local workstation cluster over a period of one year.

Interestingly, we notice in Table~\ref{tab:protIII} that the
pressure, the particle energy and the Lam\'e coefficient $\lambda$
increase  systematically in small boxes. This is shown
specifically for $\lambda$ in Fig.~\ref{figlamu}(b). These
finite-size effects indicate correlations on a similar length even
in larger systems.
We will now turn to the characterization of these correlations,
by first studying the dynamical matrix, and the quenched stresses.

\section{Contributions to the dynamical matrix: Quenched forces and spring constants}
\label{sec:Quench}
%

%\fbox{Head}

In this section we discuss various distributions and correlation functions
associated with the quenched particle positions, forces and spring constants
which contribute to the dynamical matrix (Eq.~(~\ref{eq:Dynamicmatrix})).
We attempt a characterization of the frozen in disorder visualized in Fig.~\ref{figFlink}
where the snapshots reveal strong spatially correlated fluctuations.
Specifically, we ask if
it is possible to extract a characteristic length
scale solely from the distribution of weak and rigid regions which might be a candidate
to explain the large crossover length scale $\xi$ mentioned in the Introduction.
We start by giving some additional information concerning the snapshots, turn then
to the histograms and discuss finally the fractal structure of the quenched forces.

%\fbox{Coupling constants and matrix trace.}

A timely justification for discussing quenched forces lies in the current interest
of their role in granular materials, foams and glassy colloidal systems
\cite{alexander,forcechains,chaos,liunagel,langer,mousses,claudinclement}.
It has been suggested by S.~Alexander\cite{alexander} and others that
the quenched forces might contribute to the unusual mechanical and rheological
properties in these systems.
However, to put this immediately into perspective, quenched forces are unlikely
candidates to rationalize alone the slow continuum approach discussed in Sec.~\ref{sec:Eigenmodes},
basically, since their {\em average} contribution to the dynamical matrix is weak:
$\langle r_{ij}^2 c_{ij} \rangle \gg \langle r_{ij} t_{ij}\rangle$.
This inequality is revealed in Table~\ref{tab:protIII}.
We come back to this in Sec.~\ref{sec:Eigenmodes} where we discuss
the contribution of the quenched forces to the eigenmodes.

The second and more important point we want to make here is that for
a given potential the first and the second derivative contain essentially the
same information. Both together are expressions of the same disorder generated
by the complex cooling procedure ---
which is subject to the one chosen potential and its derivatives.
It is the final positional disorder which matters, not the
individual contributions to the dynamical matrix.
If the width of the lines between the particle positions shown in the snapshots
(Fig.~\ref{figFlink}) has been chosen proportional to the interaction force rather than the spring
constant this was done mainly for {\em artistic} reasons.
Resembling snapshots can be obtained, using as a scale for the
line width either  the spring constants or  the trace
$m_{ij}=M_{ix,jx}+M_{iy,jy}$ of the dynamical submatrices. Such
pictures are direct visualizations of the dynamical matrix.
%

%
%\fbox{Discussion of histograms.}

\begin{figure}
\centerline{ 
\epsfig{file=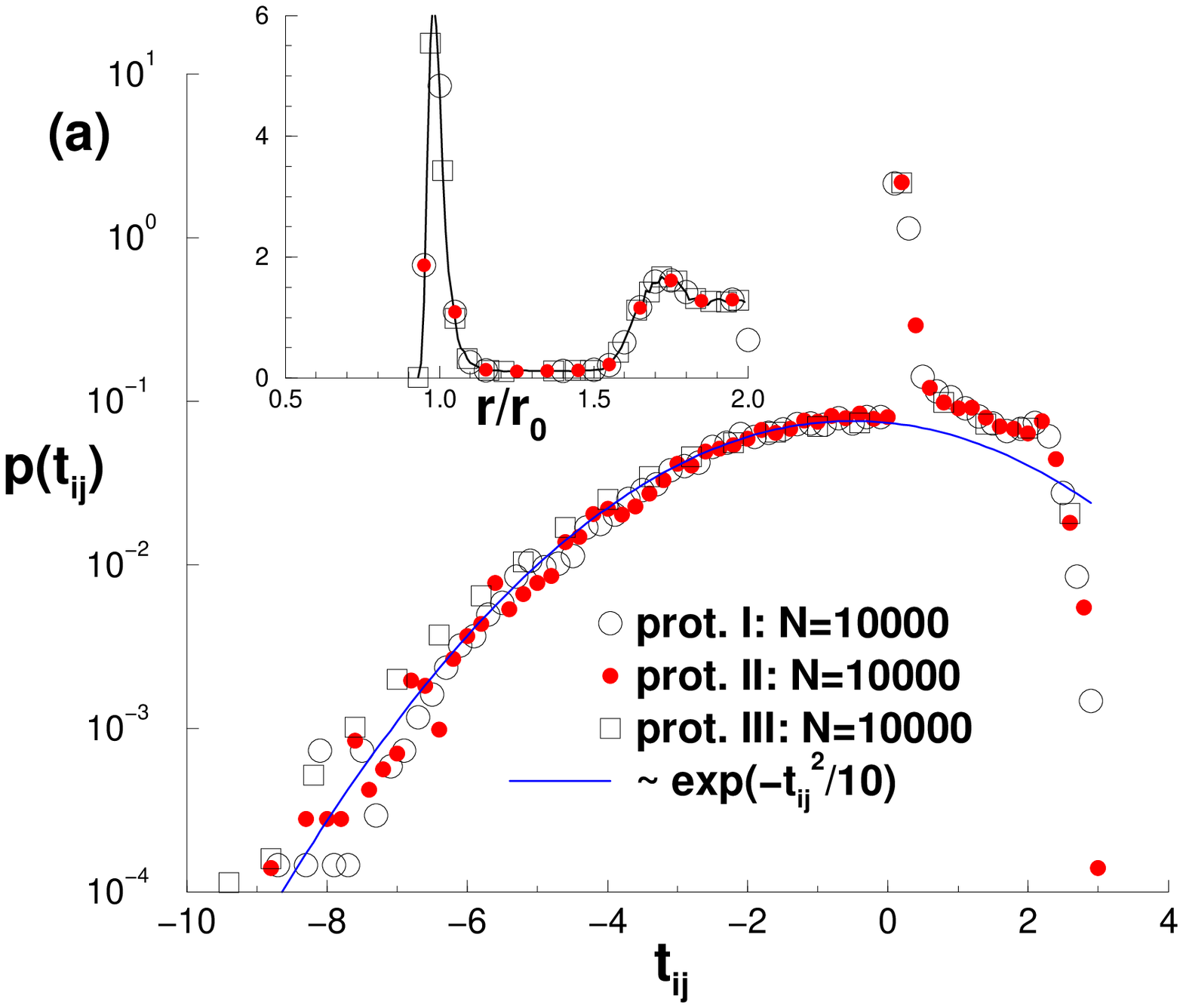,width=80mm,height=80mm}
\epsfig{file=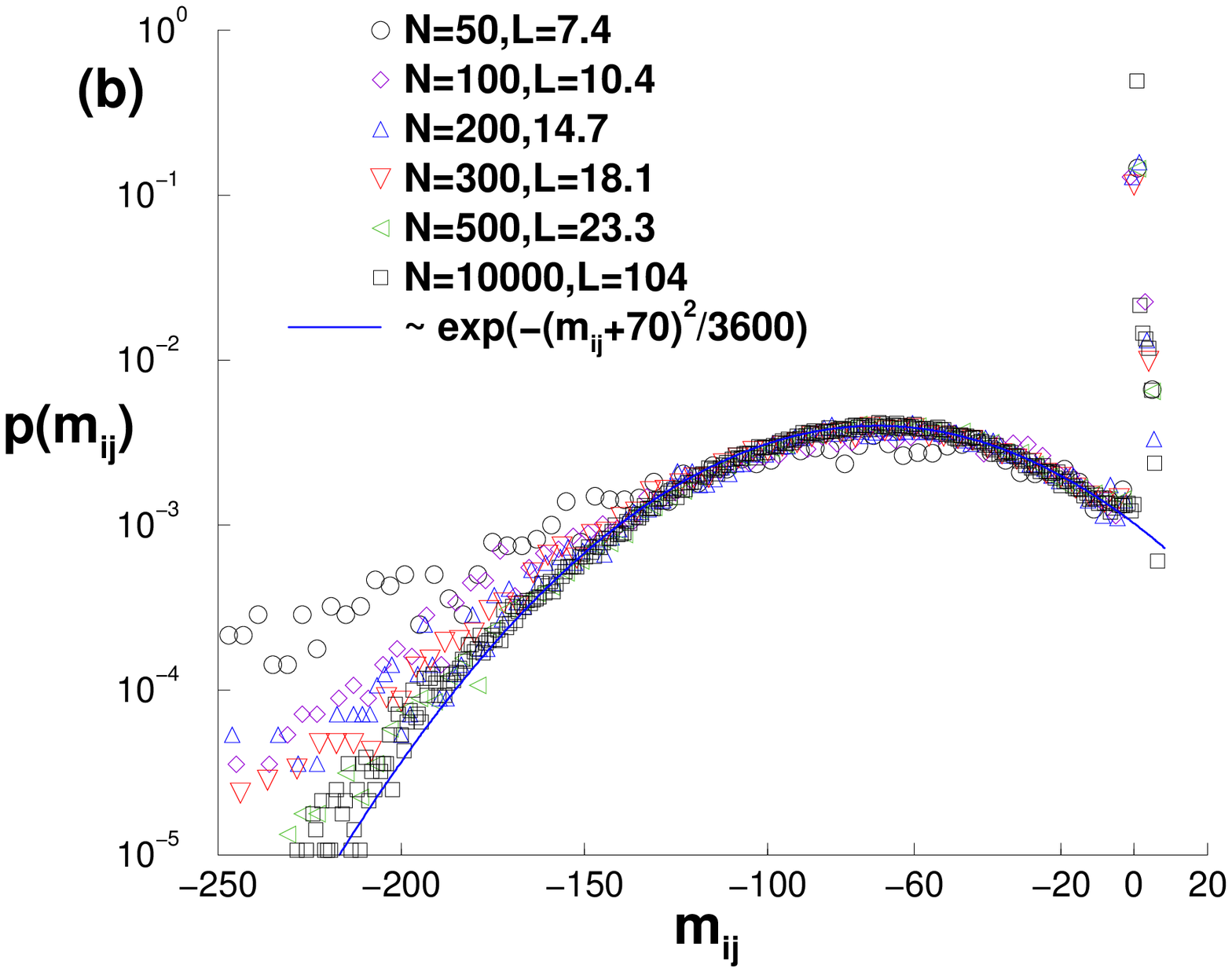,width=80mm,height=80mm}}
\vspace*{0.5cm} \caption{Histograms of contributions to
the dynamical matrix: {\bf (a)} comparison of the tension $t_{ij}$
for all three protocols at $N=10,000$. Inset: Histogram of
distances $r_{ij}/r_0$ of interacting particles.
%{\bf (b)} spring coupling constant $c_{ij}$.
{\bf (b)} trace of matrix $m_{ij}=M_{xi,xj}+M_{yi,yj}$ for systems
of various system sizes as indicated in the figure (protocol I).
This demonstrates finite-size effects in the tail of the
distributions of small systems with $L < 20$ corresponding to an
enhancement of the skeleton of very rigid contacts.
\label{fighisto}}
\end{figure}

The snapshots Fig.~\ref{figFlink} reveal that the structure of the force network is very inhomogeneous,
but isotropic on larger scale. They show also evidently that the quenched disorder is not
much affected by the different symmetries (circle or square) and quench protocols.
This corroborates the robustness of the system properties with
regard to the chosen protocol mentioned in the previous section.
This is also supported by the different histograms regrouped in Fig.~\ref{fighisto}.

Systems of the three protocols are included in first of both graphs
Fig.~\ref{fighisto}(a) where we discuss the distribution of distances
of interacting particles (inset) $r_{ij}/r_0$ and of the tensile forces $t_{ij}$
(main figure) for relatively large systems containing $N=10,000$ particles.
Comparison shows that there is essentially no dependence on quench
protocol. Note that this is only a technical point and will make
life easier when we analyse the eigenvalue spectrum of disk-shaped
aggregates.
The peak at $r_{ij}=r_0$ in the inset corresponds to the peak for $t_{ij}=0$ in
the main figure of Fig.~\ref{fighisto}(a).
Hence, most of the interactions have achieved to minimize their energy
although this is not required for global mechanical stability.
The distribution of the repulsive forces (negative tensions) is Gaussian.
Incidentally, this is quite different from the recently reported exponential force
distributions in granular matter \cite{forcechains,liunagel}.
Interestingly however, the tail of the force distribution becomes more and more exponential
upon decreasing of the system size. We do not show this here, as the same effect
is presented in the second panel (Fig.~\ref{fighisto}(b)) for the trace of the
dynamical submatrices. Various system sizes as indicated in the figure are given here
for the third protocol at $\rho=0.925$. This demonstrates that the tail of distribution
which corresponds to very rigid contacts becomes enhanced due to finite-size effect
for systems below $L\approx 20a$. A very similar figure could also be given for the
distribution of spring constants.
Note that the peak on the left side of Fig.~\ref{fighisto}(b) is due to the slow variation
of the spring constant for large distances and corresponds to distances
seen in the shoulder on the right of the histogram in the inset.

%\fbox{Discussion of fractal structure of force chains.}

\begin{figure}
\centerline{ 
\epsfig{file=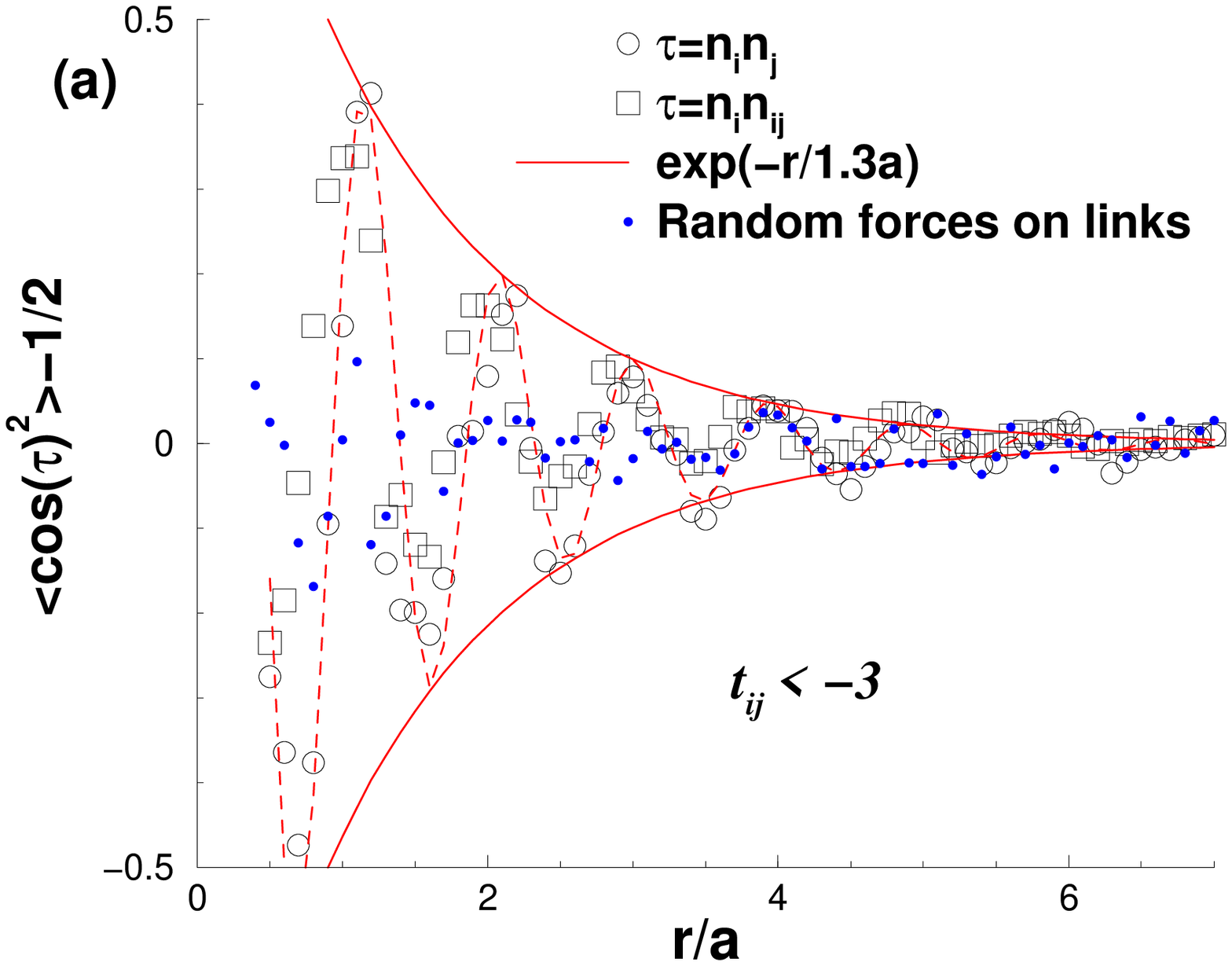,width=80mm,height=80mm}
\epsfig{file=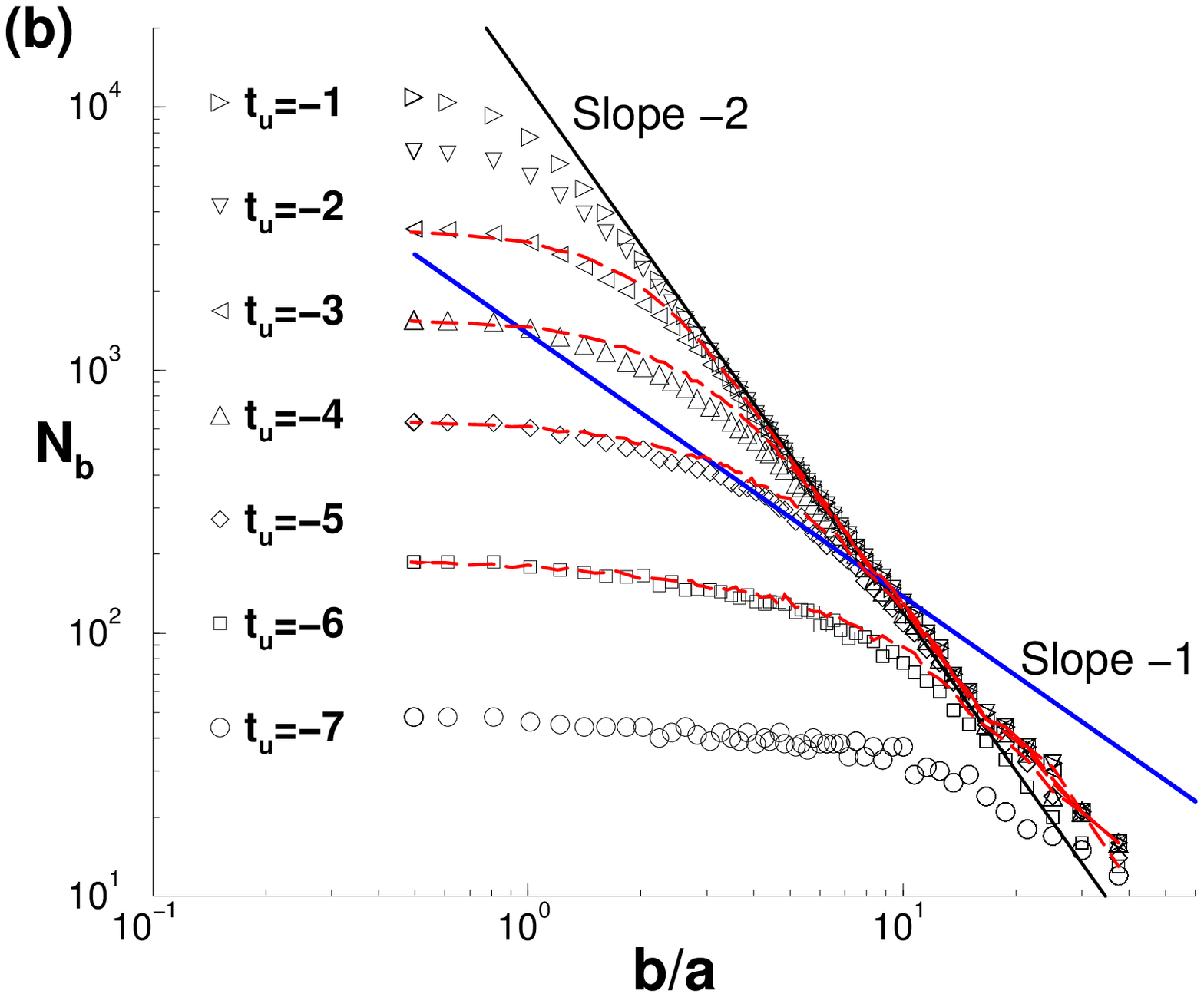,width=80mm,height=80mm}} \vspace*{0.5cm}
\caption{Correlations of frozen in forces for large
periodic systems: {\bf (a)} pair correlation $\langle cos^2(\tau)
\rangle$ {\em versus} distance $r$. Here $\tau$ denotes either the
angle between the directors $n_i$ and $n_j$ of the links $i$ and
$j$ (symbols) or the angle between the director $n_i$ of link $i$
and the direction $n_{ij}$ between the links $i$ and $j$ (lines).
The decay of the envelope is exponential with a length scale of
the order of the mean bead size, hence, this does not introduce a
new length scale. The dots correspond to an artificial force network obtained
by shuffling randomly the tensions between existing links.{\bf (b)} The number $N_b$ of boxes of size $b$
needed to cover the interactions transmitting repulsive tensions
smaller than $t_u$. The curves do not differ much from box
counting results on sets of randomly drawn links (dashed lines).
Different slopes are included for comparison. The slope -1
corresponds to linear chain like structures, the -2 slope to a
compact structure in 2D.
\label{figFcorr}}
\end{figure}

In fig.~\ref{figFcorr} we show two of several attempts to analyse
the correlations and fractal structure of force chains (or the
spring constants) in view of extracting a characteristic size. We
focus on strong repulsive forces and strong positive spring
constants. Only results from the first measurements are reported
as both yield similar results. In a first step we obtain the
network of all interactions with (negative) repulsive tensions
above some threshold $t_u$ and compute then on these sets various
correlation functions.
The functions $\langle \cos^2(\tau) \rangle$ traced in
Fig.~\ref{figFcorr}(a) attempt to put the visual impression of
linear force chains in quantitative terms.
Here $\tau$ denotes either the angle between the directors $n_i$
and $n_j$ of the links $i$ and $j$ (spheres) or the angle between
the director $n_i$ of link $i$ and the direction $n_{ij}$ between
the links $i$ and $j$ (squares). The difference between both
methods is small. The correlation dies out only after about six
oscillations. No significant difference have been found by
increasing or decreasing the threshold $t_u$. The decay of the
envelope is exponential with a length scale of the order of the
mean bead size. Hence, while rigid regions seems to be spatially
correlated,  this does not, apparently, introduce a new length
scale. Note that the situation is similar in granular materials
\cite{radjai}.

The second part of Fig.~\ref{figFcorr} shows a direct
attempt to elucidate the fractal structure of the network of
quenched forces by means of the standard box counting technique.
Here we count the number of square boxes of linear size $b$ needed to cover all the
links between beads carrying a repulsive force smaller than $t_u$.
For small $b$ where every box contains only one link the differential
fractal dimension is zero. In the opposite limit where the boxes are much
larger than the average distance between links, the differential fractal dimension
must equal the spatial dimension, i.e. $d_f=2$.
The power law slope $d_f=1$ in between both limits indicates the typical distance
where the contact network with $t_{ij}<t_u$ has a linear chain like structure.
For example, we find a tangent with slope $d_f=1$ at $b\approx 6 a$ for $t_u=-5$.
This distance is strongly increasing if we focus on more and more rigid subnetworks
(decreasing $t_u$).
Note, however, that there is no {\em finite} $b$-window with $d_f=1$
and in a strict sense there is again no characteristic length scale
associated with linear structures.
A simple visual inspection of Fig.~\ref{figFlink} that would
suggest a characteristic length scale of the repulsive force
network much larger than the interatomic separation is apparently
incorrect and possibly caused by the natural tendency of our
brains to emphasize linear patterns \cite{howthemindworks}.

As a simple benchmark for the spatial correlation of the force
network, we have distributed links randomly. The box counting of
these uncorrelated links gives the dashed lines. Unfortunately,
these are virtually identical to the fractal dimension
characterization of the force chains in the LJ systems and it is a
very tiny difference (at least in this characterization) which is
related to the spatial correlations. Obviously, this does not mean
the forces are uniformly distributed as clearly shown by the
snapshots and by the large number of oscillations in the
correlation function.

%\fbox{Summary.}

In summary, we provide evidence for a strong dispersion in the
dynamical matrix, i.e. in the {\it local} elastic properties of
our systems. Although linear chain like structures exist
apparently, this does {\em not} give rise to an additional
characteristic length
--- defined as the screening length in an exponential decay  ---
much larger than the mean particle distance.

\section{Elongation and Shear: Elastic moduli and
 non-affine displacement field}\label{sec:Nonaffine}

%\fbox{A crucial test of classical elasticity}

A direct way of illustrating the failure of classical elasticity at
small scales is to investigate the displacement field in a large,
deformed sample. If the system is uniformly strained at large
scales, e.g. by compressing or shearing a rectangular simulation box,
classical elasticity implies that the strain is uniform at all scales,
so that the atomic displacement field should be affine with respect
to the macroscopic box deformation. If this is true Eqns.~(~\ref{eq:lambdamuaffine})
should provide reliable estimates of the Lam\'e coefficients.
Since the latter can be independently measured in a computer experiment
from the generated stress differencies using Hooke's law (Eqns.~(~\ref{eq:Hooke}))
this provides a direct and crucial test for the affinity assumption.

%\fbox{How the numerical deformation was performed}

Our numerical test proceeds in three steps:
\begin{itemize}
\item
Each initial configuration is given a first quick CG quench with
{\em quadruple} precision to have a more precise estimate for the
unstrained reference system. This is necessary since the stress
differences we compute are very small for strains in the linear
response regime and the numerical accuracy of double precision
simulation runs turns out to be insufficient.
\item
Second, we perform an affine strain of order $\epsilon$ of both the box shape and
the particle coordinates.
We consider both an elongation in $x$-direction
($L_x \rightarrow L_x(1+\epsilon)$,$r_x \rightarrow r_x(1+\epsilon)$)
and a pure shear using Lees-Edwards boundary conditions \cite{AT}
together with $r_x \rightarrow r_x + r_y \epsilon$.
(We refer here to the principal box particle coordinates $(r_x,r_y)$
of the periodic configurations.)
\item
Finally, we quench with CG the configurations into the local
minima while maintaining the strain at the boundaries.  For given
boundary conditions in the linear regime the solution must be
unique.
\end{itemize}
The differences of particle positions, forces and total energies
computed at each step are recorded. We stress that  this procedure
is technically not trivial and that great care is needed to
measure physically sound properties.

We have systematically varied $\epsilon$ over several orders of magnitude
from $\epsilon=1$ down to $\epsilon=10^{-9}$ (not shown).
The induced displacement field is reversible {\and} linear in the amplitude
of the imposed initial strain for an initial strain in a window $10^{-4} \le \epsilon \le 0.1$
which decreases (for unknown reasons) somewhat with system size.
Obviously, for higher strains
the linearity of Eq.~(~\ref{eq:Hooke}) must eventually break down.
The departure from linearity below the given strain window is due
to numerical accuracy.

\begin{figure}
\centerline{
\epsfig{file=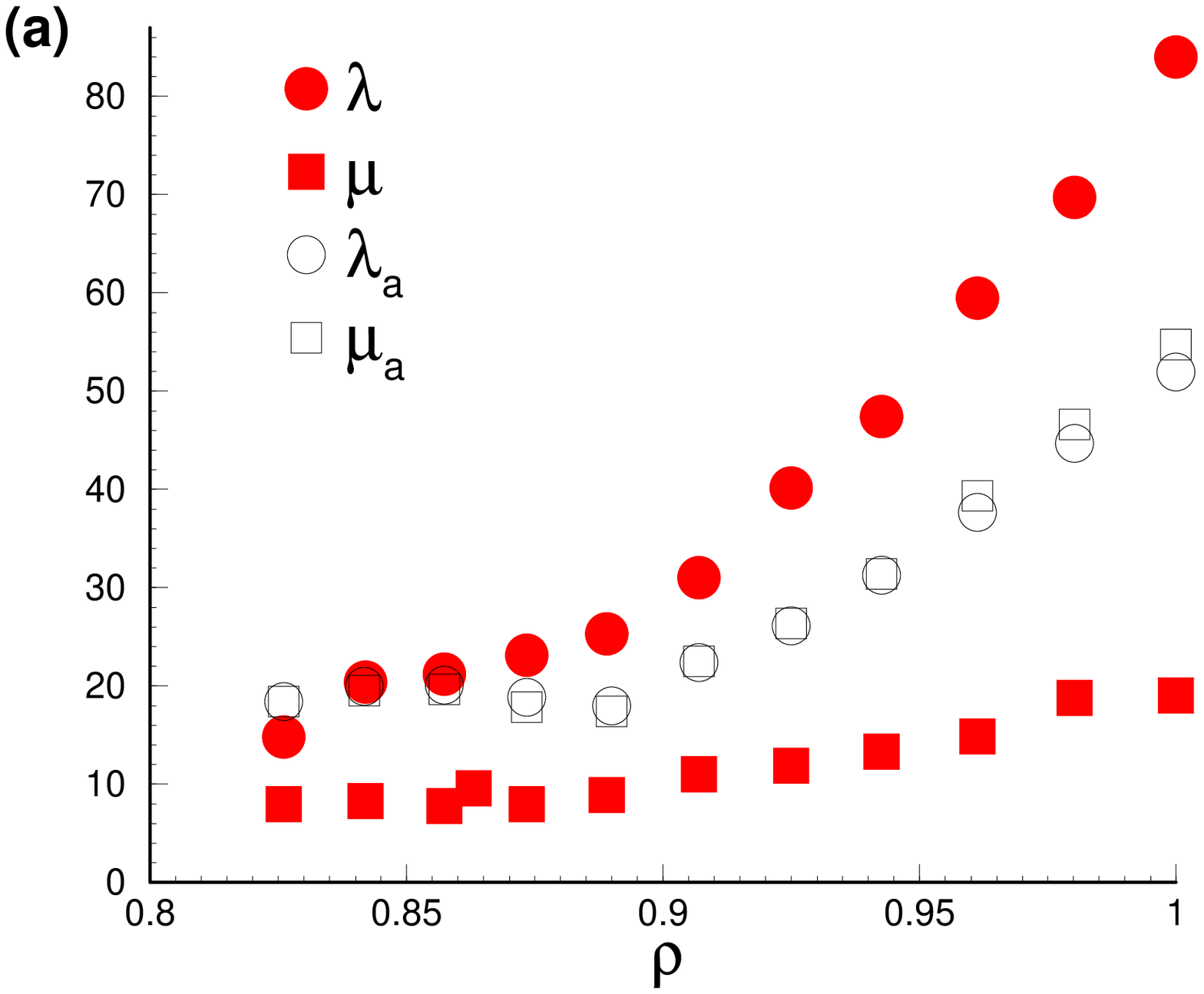,width=80mm,height=80mm}
\epsfig{file=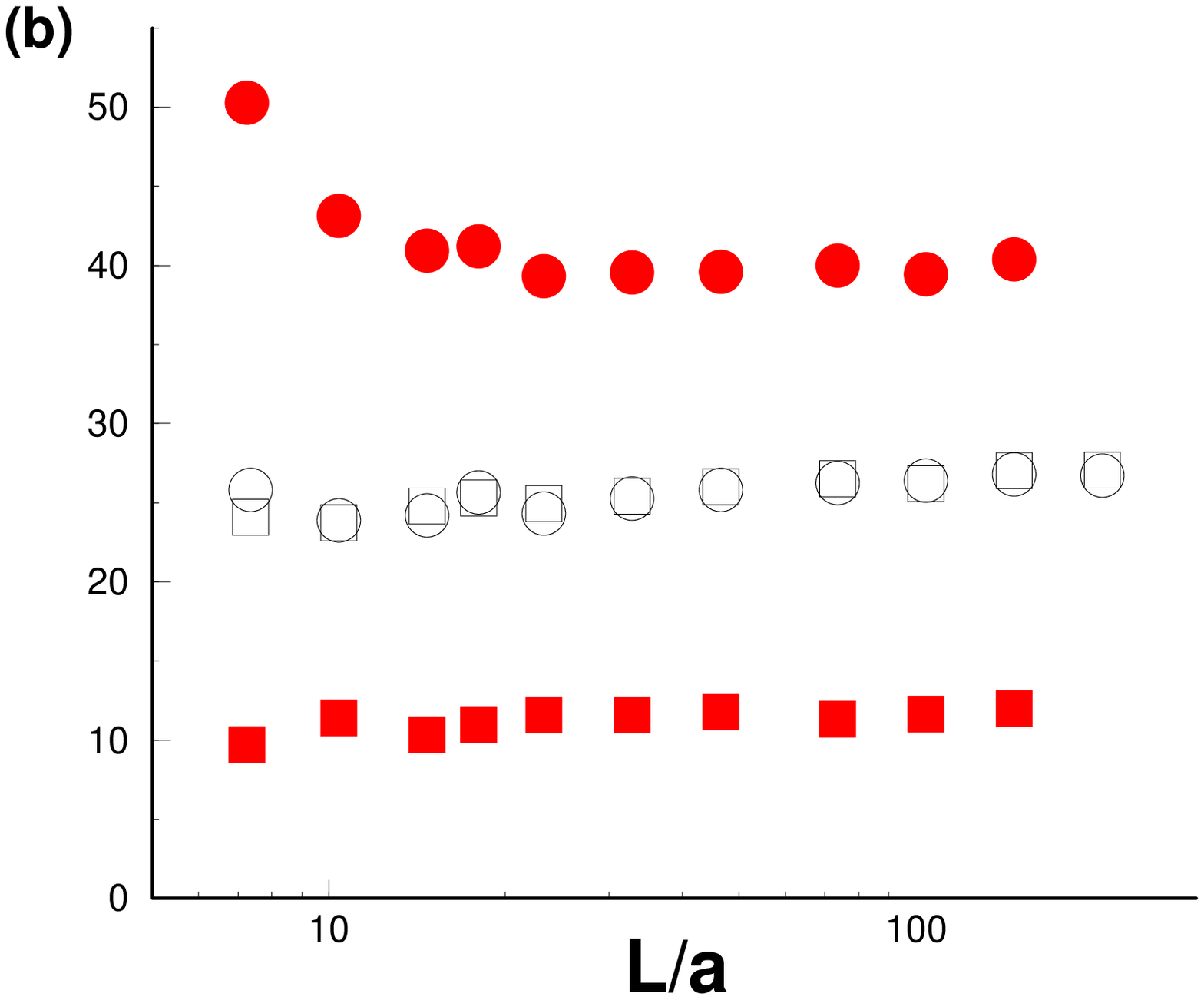,width=80mm,height=80mm}}
\vspace*{0.5cm}
\caption{
Lam\'e coefficients $\lambda$ and $\mu$
obtained for periodic bulk systems:
{\bf (a)} density variation for $N=10,000$,
{\bf (b)} size variation for $\rho=0.925$.
Full symbols correspond to the direct measurement using Hooke's law,
open symbols are obtained using eqns.~\protect\ref{eq:lambdamuaffine}.
Naturally, all Lam\'e coefficients rise with density as
the number and strength of pair interactions increases.
Note that while $\mu$ remains more or less constant $\lambda$
increases with decreasing $L$ in a similar way as pressure and
particle energy (tab.~\ref{tab:protIII}).
\label{figlamu}}
\end{figure}

Using Eq.~(~\ref{eq:Hooke}) for the stress tensor averaged on the whole system
and the averaged strain field, we obtain the true macroscopic Lamé coefficients
$\lambda$ and $\mu$. Within the given strain window the measured $\lambda$
and $\mu$ coefficients are strain independent, thus confirming the linearity of the elastic response in this range.
The values of $\mu$ and $\lambda$ are presented in
Fig.~\ref{figlamu} where we have compared them with the affine
field predictions. (Note that the Poisson ratio $\nu\approx 2/3$
is larger than $1/2$ which is permissible in a 2D system as
clarified in Sec.~\ref{sec:Theory}.)
The coefficients relying on a negligible non-affine field (open
symbols, obtained from Eq.~(~\ref{eq:lambdamuaffine})) differ by
a factor as large as two from the true ones.
Clearly, a calculation taking into account the
non-affine character of the displacements is necessary
for disordered systems.
Indeed, it is simple to work out from the values given in the
figures that, in a simple elongation, a finite energy fraction
$(\lambda_a+2\mu_a)/(\lambda+2\mu)-1 \approx 1/4$ of the total
strain can be recovered from the non-affine displacements.  In a
pure shear, the energy fraction $\mu_a/\mu \approx 1/2$ is even
larger; only the compressibility $k=\lambda + \mu$ remains
unchanged.
Hence, in quantitative terms the non-affinity of the atomic displacements
is not a negligible effect.
%

%\fbox{Non-affine displacement field}

The non-affine component $\delta \uvec$ of the atomic displacement field in large systems
subject to an elongation in $x$ direction is illustrated in the snapshot Fig.~\ref{fignoisefield}(a). A similar snapshot holds in case of a pure shear (Fig.~\ref{fignoisefield}(b)).
In some regions,
the displacement is much larger than expected from a purely affine
transformation
(Note that even the non-affine part of the displacement field
is strictly linear in $\epsilon$ within the strain interval indicated above).
Local displacements transverse to the direction of the elongation
are allowed,  and organize coherently into vortices. The
transversal direction thus cannot be neglected, showing that the
modelling approach put forward in Ref.~\cite{leibig} is not
realistic. The crossover length $\xi$ mentioned in the
Introduction manifests itself through {\em correlated deviations}
from a purely affine displacement. Visual inspection tells us that
the sizes of the vortices and $\xi \approx 30 a$ are comparable.

%\fbox{Correlation functions}

The two correlation functions presented in Fig.~\ref{figbcorr}
confirm this visual impression. The first one shows the
correlation function $C_u(r) = \langle \delta \uvec(r) \cdot
\delta \uvec(0) \rangle$. The striking anti-correlation for $r \ll
30a$ is in agreement with the size of the vortices seen in the
displacement fields.
%and would correspond  to the ``Taylor
%length'' in fluid mechanics~\cite{turbulence}.
%
That the displacement field is indeed correlated over a similar size is
further elucidated in Fig.~\ref{figbcorr}(b). Here we consider the systematic
coarse-graining of the non-affine displacement field
\begin{equation}
\delta \ucoarse_j \equiv \frac{1}{N_j} \sum_{i \in V_j} \delta \uvec(\rvec_i) 
\end{equation}
of all $N_j$ beads contained within the square volume element $V_j$ of linear size $b$.
The mean-squared average
$U_x(b) \equiv \langle  \delta {\cal U}_{x,j} ^2  \rangle_j^{1/2}$
is plotted versus the size of the coarse-graining volume element $b$.
We have normalized the function by its value at $b=1$.
The coarse-grained field decreases very weakly for $b < 30a$
and only for much larger volume elements we find
the power law slope $-1$ expected for uncorrelated events.
As for symmetry reasons the total or mean non-affine field
vanishes we have $U_x(b\rightarrow L) \rightarrow 0$ for very large
volume elements. Apart from this trivial system size dependence
$U_x$ approaches a system size independent envelope for $b \ll L$,
as can be clearly seen from the figure.

Barely distinguishable functions have been obtained for $U_y$
(not depicted) which demonstrates the isotropy of the non-affine displacement fields
which may also be inferred straight from the snapshots and appropriately
chosen correlations functions.
Similar characterizations can be obtained from a standard
Fourier transform of  $\delta \uvec(x,y)$ and from the gradient fields defined
on the coarse-grained field $\delta \ucoarse_j$. These are again not presented.
%

%\fbox{Stress tensor}

\begin{figure}
\centerline{
\epsfig{file=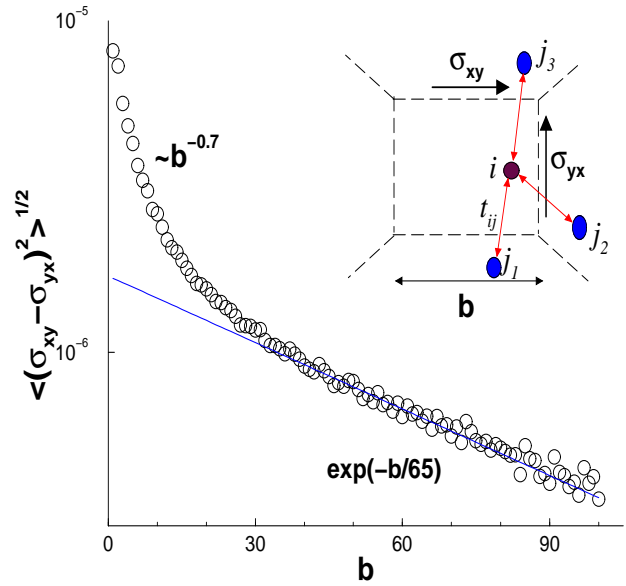,width=80mm,height=80mm}}
\vspace*{0.5cm}
\caption{
Asymmetry of the stress tensor of the forces generated in simple shear
(Lees-Edwards boundary conditions) for a system containing 10 000 particles.
We measure the ''microscopic stress`` acting on a volume element
as shown in the sketch on the right.
In the main figure we plot the average mean-squared stress difference
$\langle (\sigma_{xy}-\sigma_{yx})^2\rangle^{1/2}$ versus the linear size $b$
of the volume element. For small box sizes $b<30 a$ we evidence power law behavior
which crosses to an exponential decay at large volume sizes.
\label{sigmaxy}}
\end{figure}

In the rest of this section,  we consider the local stresses
generated by the applied macroscopic deformation. We show in
Fig.~\ref{sigmaxy} the variance $\langle
(\sigma_{xy}-\sigma_{yx})^2\rangle$ averaged on various boxes of
size $b$. Here, the stress tensor $\sigma_{\alpha\beta}$ has been
defined, as in classical mechanics, as the average force exerted
in the $\alpha $ direction through the side perpendicular to the
$\beta $ direction of the volume element of size $b$. We see
clearly in the Fig.~\ref{sigmaxy} that for a size $b < 30 a$, this
stress tensor is asymmetric, and that the asymmetry decreases
exponentially to zero for larger sizes. In agreement with the
discussion of section~\ref{sec:Theory}, this effect must be due to
the lack of invariance of the energy under translations or
rotations for volume elements of sizes $< 30 a$.

%
%\fbox{Finite-Size effects}

We note that the length scales observed in all these plots
are relatively large compared to the particle size, but are {\em not}
in the classical sense characteristic lengths appearing in the
exponential decay of a correlation function.
An important consequence of the large spatial correlations is
that calculations of Lam\'e coefficients are prone to
finite-size effects for system sizes similar and below $\xi$.
This explains qualitatively the peculiar system size dependence of the Lam\'e
coefficients reported in Fig.~\ref{figlamu}(b).

\section{Vibration modes}\label{sec:Eigenmodes}

In this section we discuss finally the eigenvalues and eigenvectors
of the different systems we have generated.
For each configuration the lowest ($p \le 1000$) vibration
eigenfrequencies and eigenvectors have been determined using the
version of the Lanczos method implemented in the PARPACK numerical
package \cite{arpack}. As stressed in the introduction we
concentrate on the lowest end of the vibrational spectrum, since
this is the part that corresponds to the largest wavelengths for
the vibrations.

We continue and finish first the discussion of disk-shaped aggregates
which started with the snapshot Fig.~\ref{figFlink}(a).
This being done we focus more extensively on the simpler periodic glassy systems (Protocol III) and discuss
subsequently their eigenfrequencies and eigenvectors. This allows us to pay
attention to central questions of strongly disordered elastic materials
without being sidetracked by additional physics at cluster boundaries
(ellipticity, radial variation of material properties etc.).

\subsection{Eigenmodes of disk-shaped aggregates}

%\fbox{$\omega$ versus $p$}

\begin{figure}
\centerline{
\epsfig{file=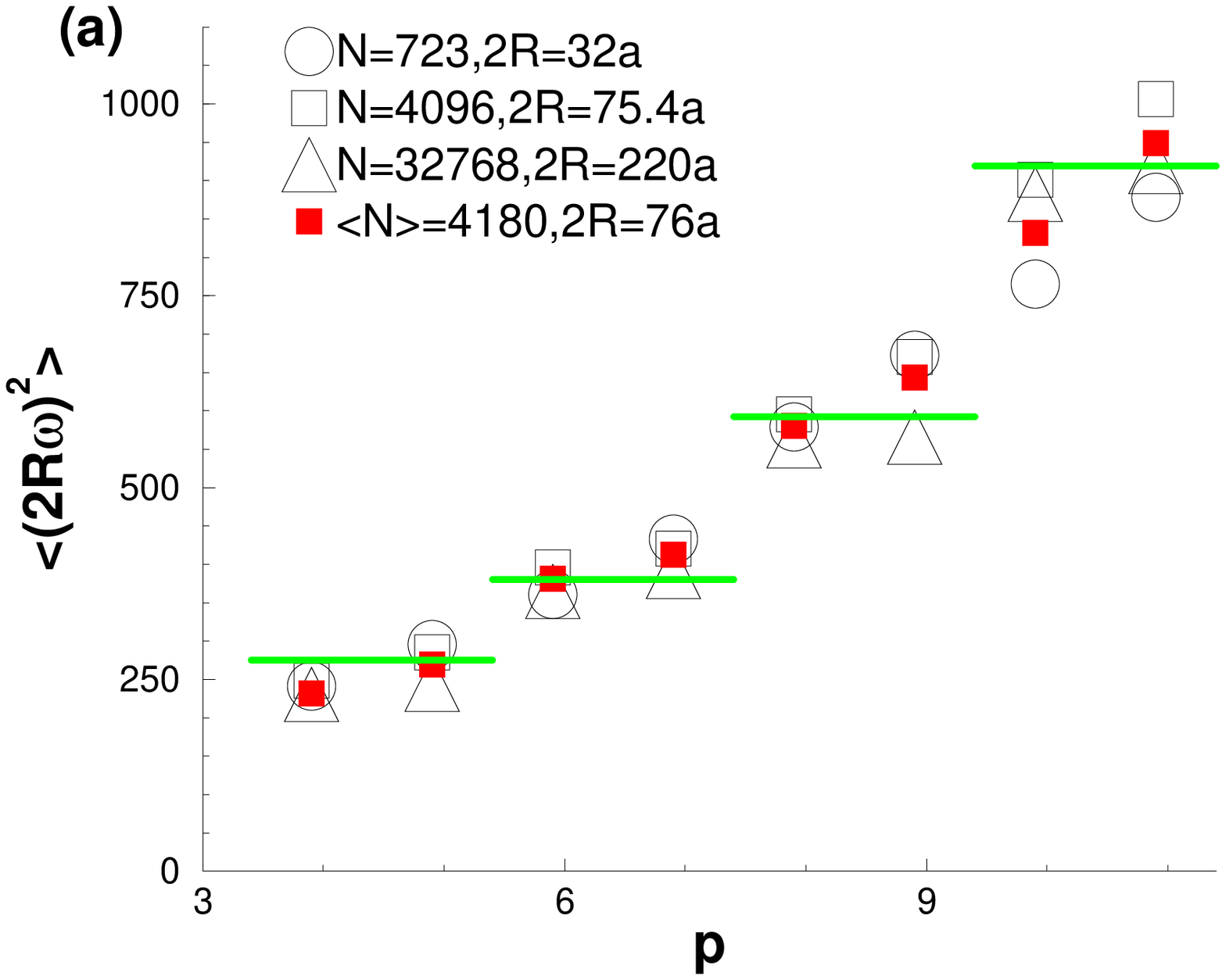,width=80mm,height=80mm}
\epsfig{file=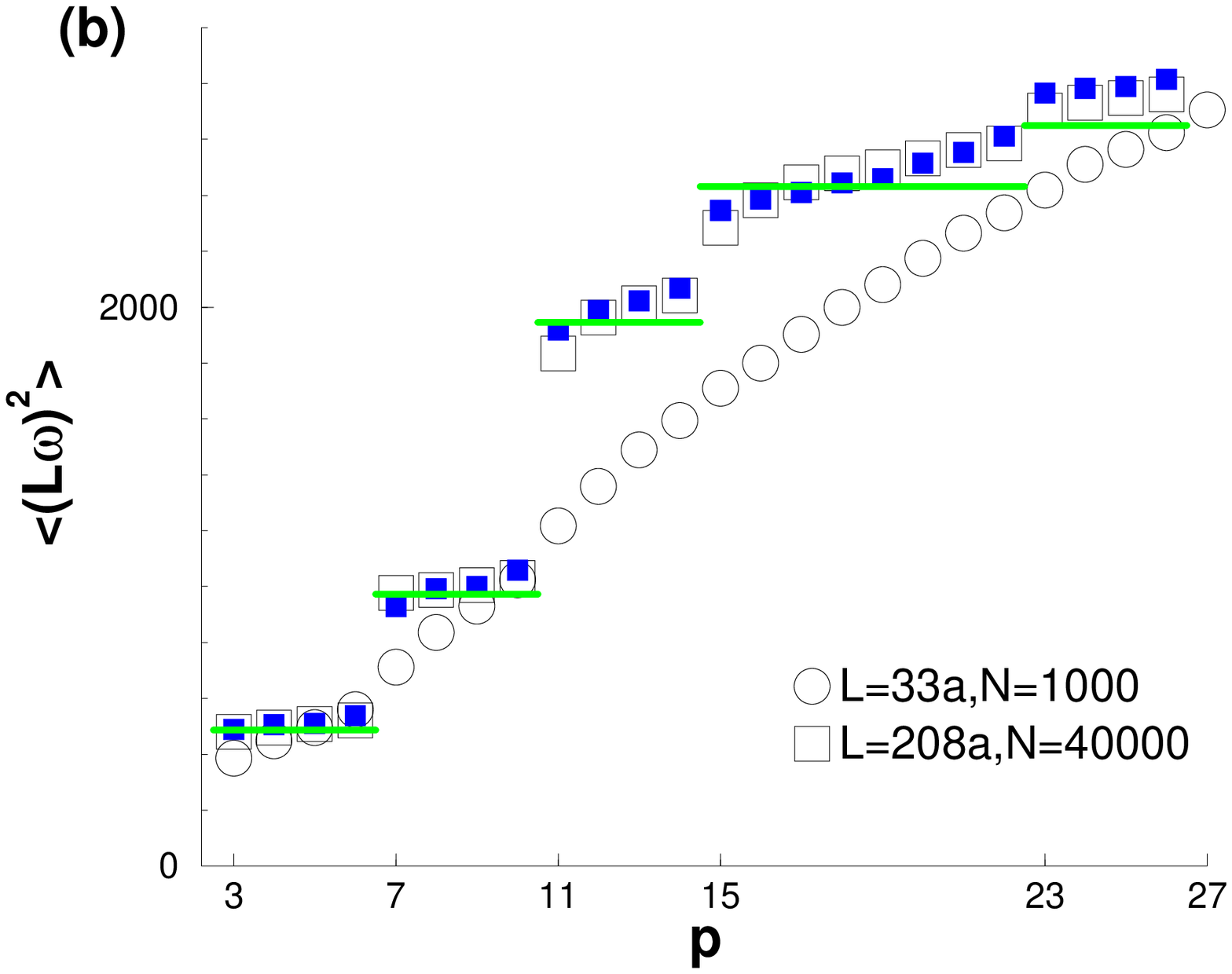,width=80mm,height=80mm}}
\vspace*{0.5cm}
\caption{
Comparison of the first non-trivial eigenvalues
for {\bf (a)} disk-shaped clusters and {\bf (b)} bulk systems
confirming the predicted degeneracies for sufficiently large samples.
In small systems, the degeneracy is lifted.
The frequencies are rescaled with the system size (as indicated)
and compared with the theoretical predictions (horizontal lines).
The open symbols in {\bf (a)} correspond to the slow quench (Protocol I),
the full symbols to a rapid quench (Protocol II)
showing that the results of both protocols become similar for large disks.
The open and full squares in {\bf (b)} correspond to two different configurations
with the same parameters.
\label{figEW}}
\end{figure}

The first non-trivial eigenvalues ($4 \le p \le 11$) for the two protocols
for disk-shaped clusters are shown in Fig.~\ref{figEW}(a).
The first three eigenmodes have vanishing eigenfrequencies because
of two translational and one rotational invariance.
Aggregates of different sizes are presented as indicated in the figure.
The frequencies are rescaled with the disk diameter $2R$ as suggested
by dimensional considerations or continuum theory.
This scaling is roughly successful for all systems included.
For the smaller systems (e.g., for the example with $N=732$ given)
$\omega$ do not present the degeneracies of the continuum theory.
If we increase the system size steps appear and the eigenfrequencies
start to regroup in pairs of two following the continuum prediction.
This is well verified for the largest disk we have created containing
$N=32,768$ beads.
The horizontal lines are comparisons with continuum theory with
appropriate density and where
the Lam\'e coefficients  (and, hence, the sound velocities)
have been taken accordingly
from Table~\ref{tab:protIII}.
The comparison of the two protocols (I and II) for $N=4096$ shows that,
perhaps surprisingly, the quench protocol does not matter much
if only the clusters are large enough and the excentricity sufficiently
weak. Note that this is definitely not true for small disks where
the excentricity matters.

%\fbox{Eigenvectors}

\begin{figure}
\centerline{\epsfig{file=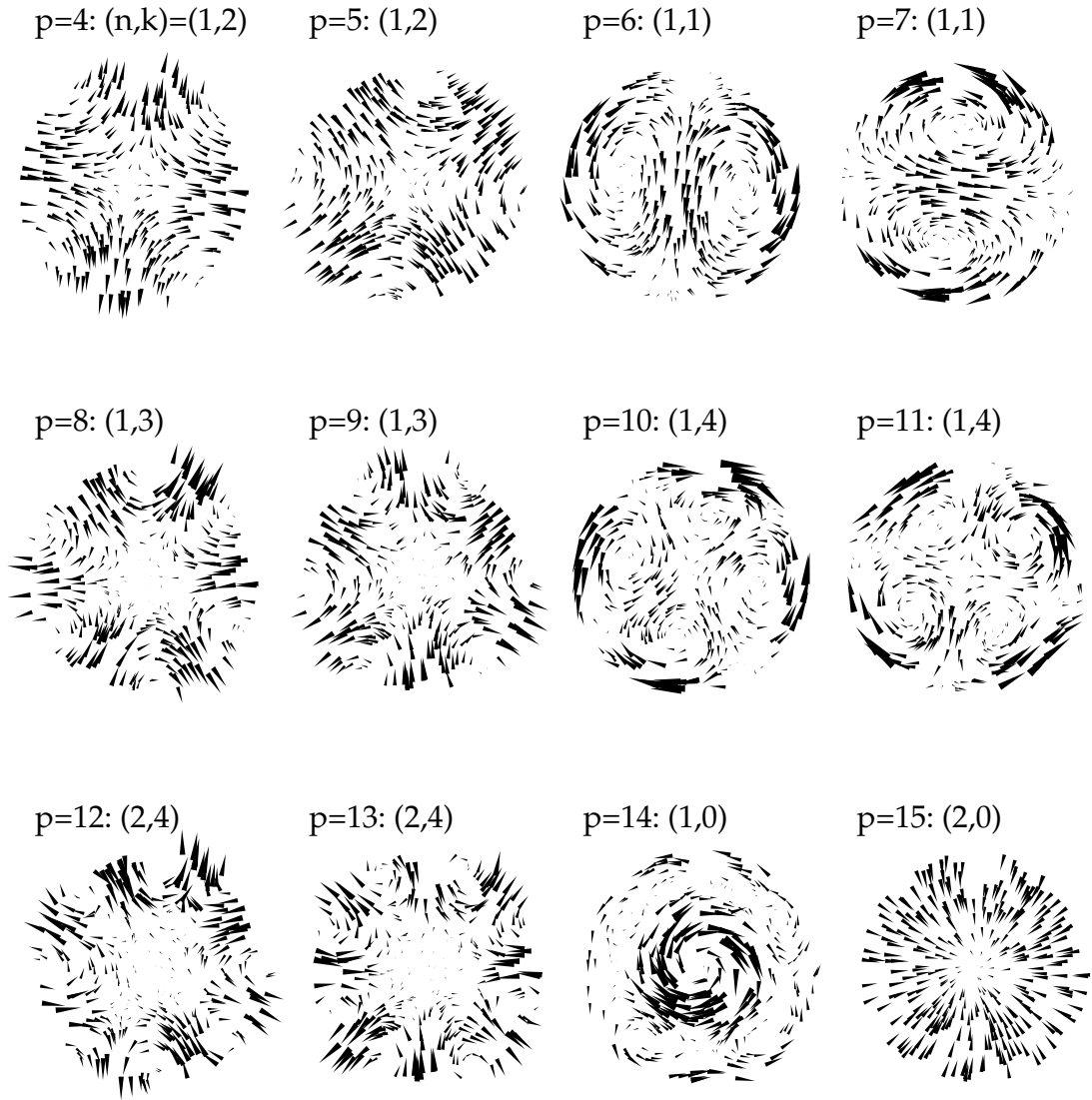,width=150mm,height=150mm}}
\vspace*{0.5cm}
\caption{
First non-trivial eigenvectors (from $p=4$ up to $p=15$) of an aggregate
of diameter $2R=120a$ containing $N=10,000$ particles (protocol I). We have indicated
the running index $p$ and the pair
of qantum numbers $(n,k)$. The two-fold degeneracy for $k>0$ is obvious.
Note that each pair is rotated by a polar angle $\Delta\theta=\pi/2k$.
The last mode represented is an axialsymmetric  `breathing mode' ($k=0$).
\label{figEVagg}}
\end{figure}

A schematic representation of the displacement fields for the lowest eigenmodes of a
large disk-shaped cluster ($R=60$) is given in Fig.~\ref{figEVagg}.
For such large cluster the displacement fields essentially conform to standard
elastic behavior. The two quantum numbers $n$ and $k$ are indicated for each mode.
The snapshots demonstrate nicely the two-fold degeneracy for all the modes
which are not axisymmetric (the first ten fields).
We confirm that corresponding pairs are turned with respect to each other by
an angle $\Delta\theta=\pi/2k$, e.g. for the first non-trivial modes $p=4$ and $p=5$
where $k=2$ by an angle $\Delta\theta=45^0$.
The disk wobbles in these modes between two perpendicular slightly elliptic shapes.
The numerics returns arbitrary linear superpositions of
eigenvectors associated with the same eigenvalue. As we have at
most a two-fold degeneracy this does, however, only amount to an
arbitrary and common rotation of the pairs of snapshots given.
(The situation is more complex for periodic systems, see below.)
Examples for the not degenerated axially symmetric case ($k=0$)
are the `watch spring' $p=14$ and the `breathing mode' $p=15$. The
last mode is typically excited in Raman spectroscopy
\cite{vallee}.
%

%\fbox{Finite-Size analysis}

The next step is now to characterize the continuum approach
as a function of system size. Our analysis presented in
Fig.~\ref{figEWfs}(a) is similar to the finite-size studies
of phase transitions and critical phenomena.
The rescaled eigenfrequencies for given $p$ are plotted versus the rescaled
inverse system size $\xi/2\langle R\rangle$ in such a way that the
vertical axis $\Omega^2 = \langle (\omega R/\pi c)^2 \rangle$
should become independent of the cluster properties (size, density, Lam\'e coefficients)
in the limit of large systems.
$\xi$ is the cross-over length defined in the previous section. We have taken $\xi \equiv 30a$.
The horizontal lines correspond to the
continuum predictions for quantum numbers $(n,k)$ as indicated.
Both axes are dimensionless.
For the $p$ given in the figure only transversal modes
are expected and, hence, we have used the transversal sound
velocity $c=c_T(\rho)$ everywhere.
For $p < 14$ all modes should be two-fold degenerated and one expects
even (full symbols) and odd (open symbols) modes to regroup in pairs. This is born
out for the larger systems; for $2\langle R \rangle \gg \xi$ the
lowest frequencies even match {\em quantitatively} the predictions.
There is no adjustable parameter left for the vertical scale.
The crossover at $\xi/2\langle R \rangle \approx 1$ for the smallest eigenfrequencies
justifies the (somewhat arbitrary) choice of the numerical value of $\xi$.
Interestingly, the continuum limit is approached in a {\em non-monotonous} fashion
and very small systems vibrate at higher frequencies. One cause for this is certainly
the higher excentricity of smaller clusters. As we will see below, however,
additional and more fundamental physics plays also a role.

Every data point corresponds to an average over an ensemble with
identical operational parameters. The number of configurations in
every ensemble have been chosen such that the (not indicated)
error bar is of the order of the symbol size. The dispersion of an
individual measurement is much larger, however, for smaller
systems and of the order of the frequency difference between
subsequent modes $\omega(p+1)^2-\omega(p)^2$. As the problem seems
to be strongly self-averaging, as one expects,  the dispersion
between different representations of an ensemble goes strongly
down with system size.
Note that the diameters $2R$ and the densities $\rho$ of each
configuration in an ensemble vary somewhat and we have used in the
averaging procedure the sound velocities associated with every
specific sample density. This was done by means of interpolating
the numerical values of the  Lam\'e coefficients shown in
Fig.~\ref{figlamu}(a). The dispersion of sound velocities within
an ensemble is, however, relatively weak even though the Lam\'e
coefficients depend strongly on density.

We have presented in the figure data from the first  protocol of
elaboration. Data from the second protocol looks quite similar.
For small systems there are differences probably due to the higher
eccentricity as mentioned above. As some of the results presented
here, such as the non-monotonous variation, could be due to
spurious surface and eccentricity effects in circular systems, we
now turn our attention for the rest of this section to the
periodic glassy systems.

\subsection{Eigenfrequencies of periodic bulk systems}

%\fbox{$\omega$ versus $p$}

Raw data for eigenfrequencies for systems generated following the third protocol
are given in Fig.~\ref{figEW}(b) for two examples at $\rho=0.925$.
As there is no rotational invariance in a periodic
box only the first two modes $p=1$ and $p=2$ vanish.
The vibration frequencies  do not display the degeneracies of the
continuum in the smaller system with $L=32.9$ (spheres). It
appears that the finite-size effects are much more pronounced in
the eigenfrequencies compared to the weak effects discussed in
Fig.~\ref{fighisto}(b) on the stiffness.
In contrast to small systems, the  degeneracy steps are clearly
visible for the largest configurations (square symbols) we have
sampled with $N=40,000$. The quantitative agreement with continuum
prediction is then satisfactory and deteriorates only slightly
with increasing $p$, i.e. with decreasing wavelength $\lambda(p)$.
Two configurations have been obtained in the latter case (open and
full symbols). Interestingly, the self-averaging is such that both
are barely different, even where they depart from the classical
theory.
%

%\fbox{Finite-Size scaling}

\begin{figure}
\centerline{
\epsfig{file=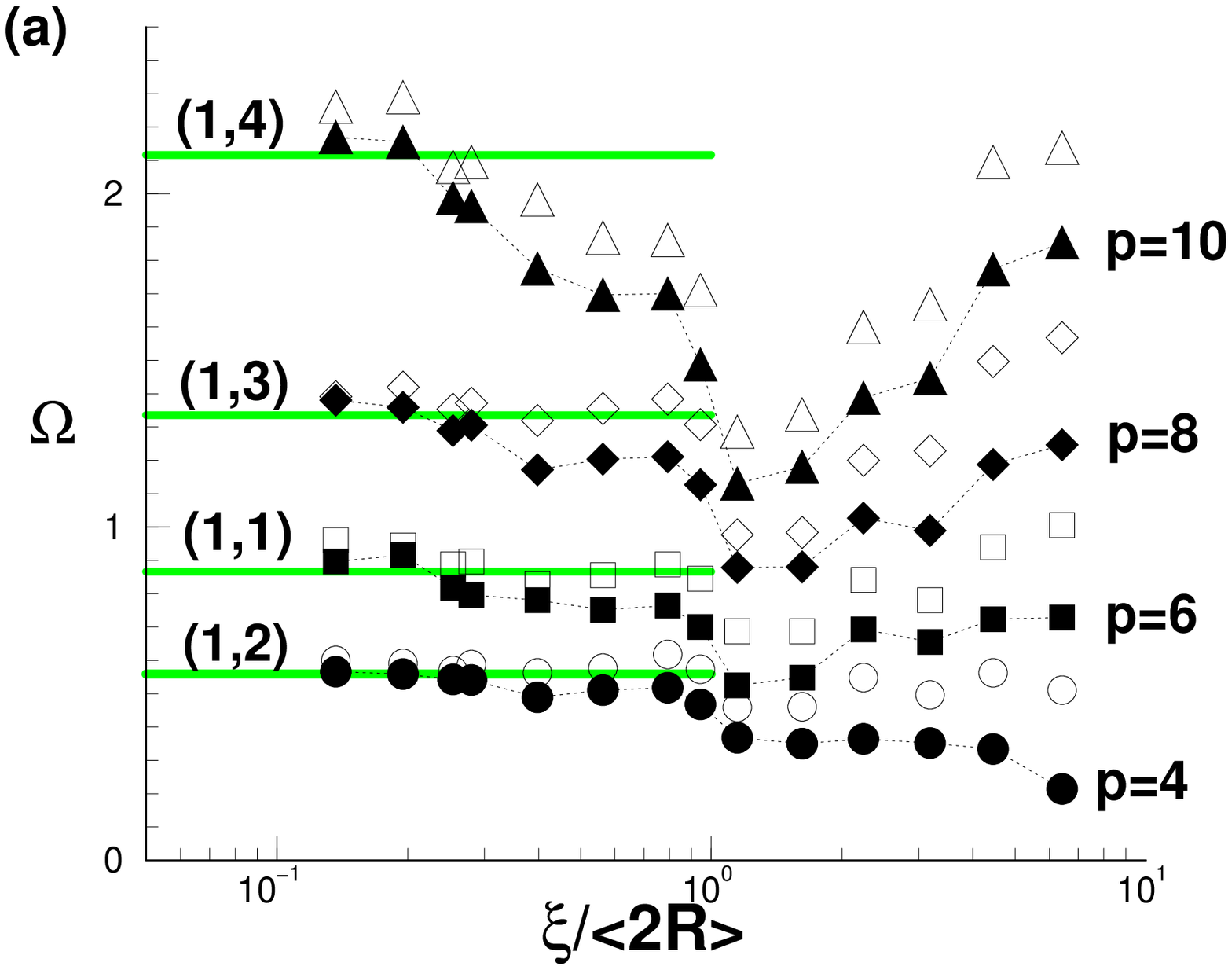,width=80mm,height=80mm}
\epsfig{file=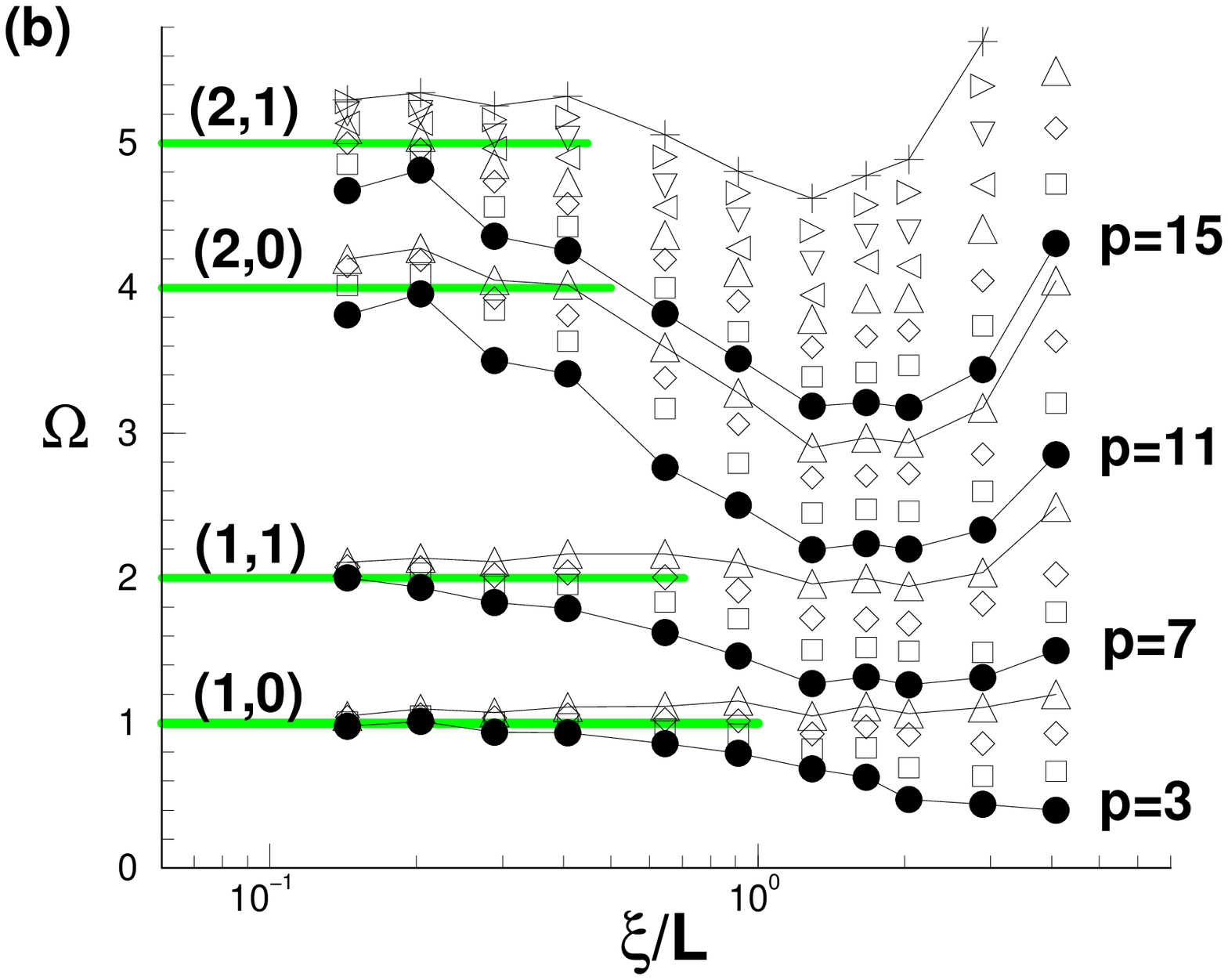,width=80mm,height=80mm}}
\vspace*{0.5cm}
\caption{
Finite-size scaling of reduced eigenfrequencies
$\Omega(p,L)=\langle (\omega(p,L) L/ 2\pi c_T)^2 \rangle$ {\em versus} $\xi/L$
with $\xi\equiv 30 a$ :
{\bf (a)} disk-shaped aggregates from protocol I with $L=2R$,
{\bf (b)} periodic bulk systems at $\rho=0.925$ (protocol III).
The mode index $p$ increases from bottom to top, some of the $p$
are specifically given (full symbols).
In both cases the degeneracy is systematically lifted for small systems
and the continuum prediction (given by the horizontal lines)
is approached non-monotoneously.
The pairs of quantum numbers $(n,k)$ and $(n,m)$ associated with the
predictions are indicated in figure {\bf (a)} and {\bf (b)} respectively.
\label{figEWfs}}
\end{figure}

Fig.~\ref{figEWfs}(b) shows the eigenfrequencies for bulk systems
as a function of box size $L$ in analogy to the characterization
presented above for disks. The horizontal axis is now $\xi/L$,
the vertical axis $\Omega^2=\langle (\omega L/ 2\pi c_T)^2 \rangle$.
While the continuum approach is somewhat smoother in the bulk case
than for the disks essentially both sets of data shown in
Fig.~\ref{figEWfs} carry the same message: They indicate that for
system sizes $L$ below $\xi$ the predictions of continuum
elasticity become erroneous even for the smallest eigenmodes. The
classical degeneracy of the vibration eigenfrequencies is lifted
and the resulting density of states, becomes a continuous
function. The approach of the elastic limit is again
non-monotonous. This is not related to the dispersion due to the
discreteness of an atomic model, which would result into a
monotonous approach to the elastic limit, as can be easily checked
on one dimensional models. As we have no surface effects here the
effect  must be due the frozen-in disorder. Obviously, the physics
at play should also be relevant for the disk-shaped clusters.

%\fbox{Quenched stresses.}

As a sideline we report here briefly that we have also
investigated the role of the quenched stresses on the
eigenfrequency spectrum. This can be readily done by switching off
the contribution from the tensions in the dynamical matrix
Eq.~(~\ref{eq:Dynamicmatrix}). The corresponding {\em artificial}
system appears to remain mechanically stable (i.e. all eigenvalues
remain positive),  however, it does not correspond to any
realistic interaction potential. A finite-size plot analog to the
ones shown in fig.~\ref{figEWfs} has been computed for periodic
systems. This yields a result {\em qualitatively} very similar to
the curves presented here, albeit the crossover to continuum
occurs for slightly smaller box sizes.
The point we want to make here is two-fold: On the one side
quenched forces matter if it comes to {\em quantitative}
comparisons and evaluation of a analytical model, on the other
hand, they do not generate new physics. The role of quenched
stresses is simply to maintain a local equilibrium in systems with
strong positional disorder. The results presented here, and
particularly the deviation from classical continuum theory,
appears to be due to local disorder and simple harmonic but
disordered couplings would give analogous results.

%\fbox{Wavelength scaling of eigenvalues}

\begin{figure}
\centerline{
\epsfig{file=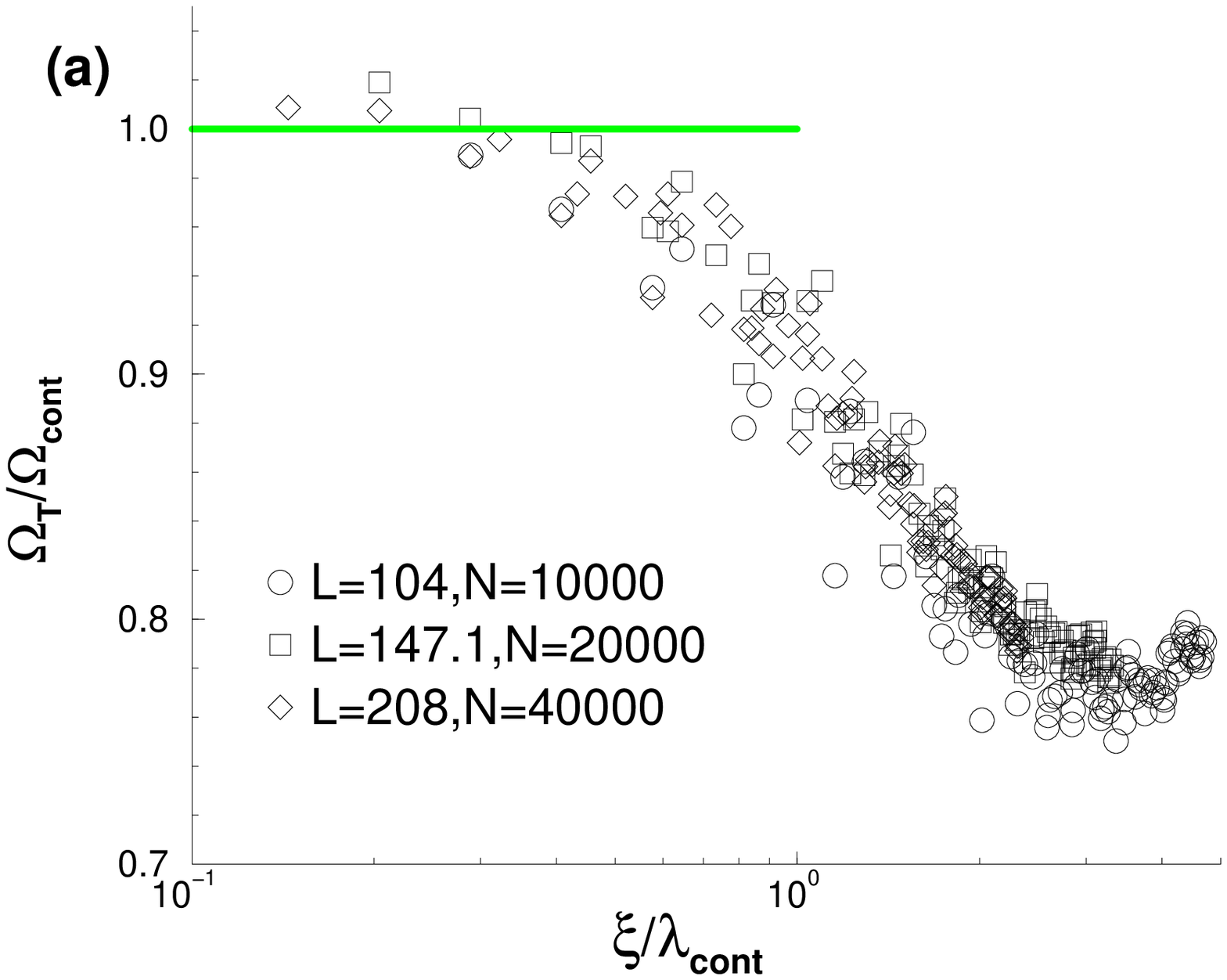,width=80mm,height=80mm}
\epsfig{file=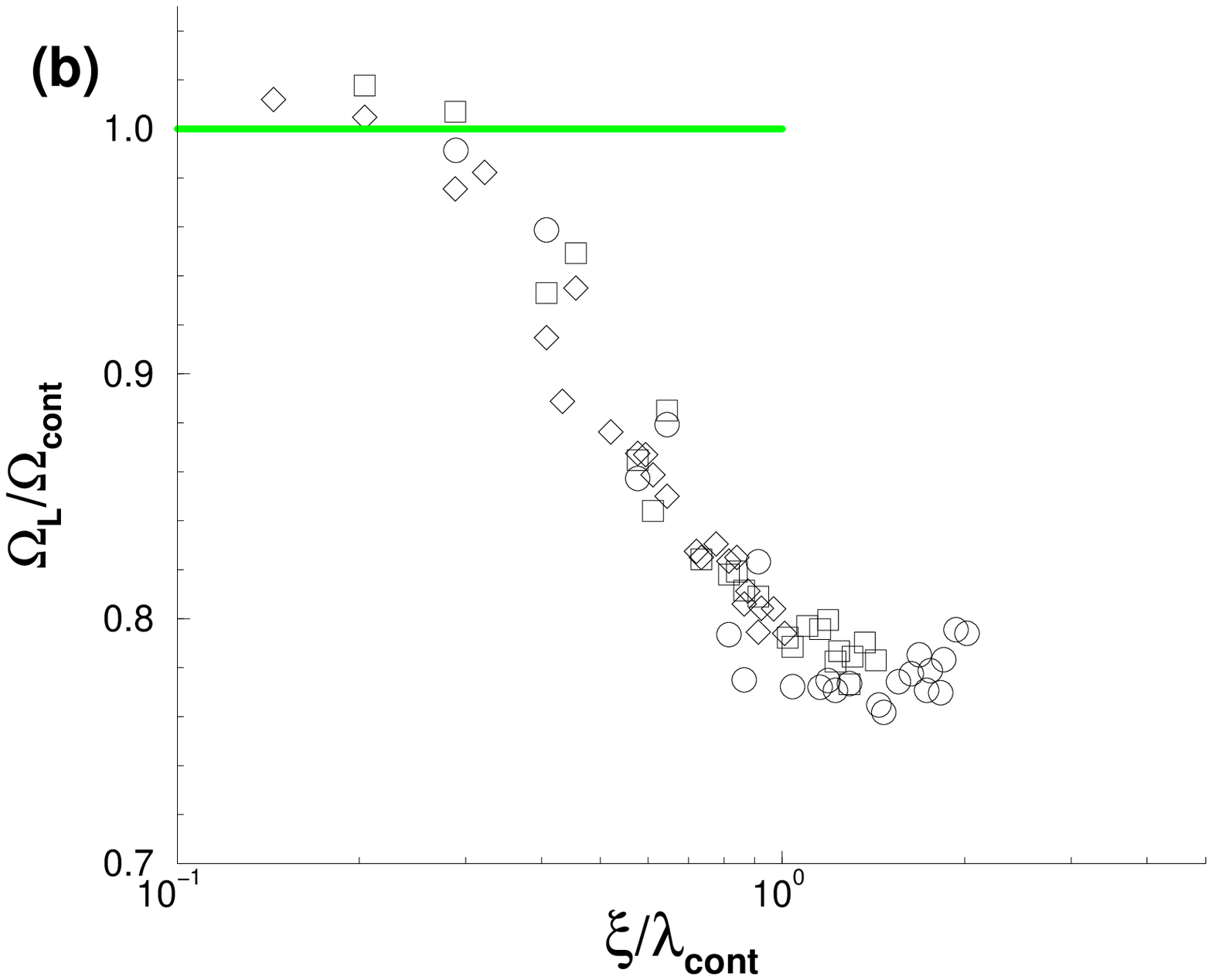,width=80mm,height=80mm}
}
\vspace*{0.5cm}
\caption{
Scaling of {\bf (a)} transverse and {\bf (b)} longitudinal modes for different box sizes as indicated.
The rescaled frequency $\Omega/\Omega_{cont} = \langle \omega^2 \rangle/\omega_{cont}^2$
%%%_{T,L}(p,L)=\langle (\frac{L\omega}{2\pi c_{T,L}})^2\rangle$
is plotted versus the inverse wave length $\xi/\lambda_{cont}$.
The theoretically expected wave length $\lambda_{cont}=L/(n^2+m^2)$
is given by the quantum numbers $(n,m)$ associated with the mode index $p$.
We have again set $\xi \equiv 30a$.
The crossover to continuum theory occurs at $\lambda \approx \xi$ for transverse modes
and at about twice as large wave lengths for longitudinal modes.
The success of both scaling plot suggests that $\xi$ is frequency independent
for sufficiently large system sizes where $\lambda(p) \ll L$, but does depend the wave type.
We regard this as the new central results of this paper.
\label{figyscal}}
\end{figure}

As in the study of critical phenomena where finite-size effects
reveal correlations at wavelengths $\lambda(L,p)$ in much larger systems
where $L \gg \lambda$ one naturally expects that the results of Fig.~\ref{figEWfs}
are also relevant for the description of higher eigenmodes
and their departure from continuum theory.
We demonstrate in Fig.~\ref{figyscal} that the
crossover to continuum is indeed characterized by the ratio
of  wavelength and the correlation length $\xi$.
The numerical eigenfrequencies $\Omega$ are rescaled by their
expectation $\Omega_{cont}(n,m)$ from continuum theory
(Eq.~(~\ref{eq:evperiodic})) and plotted versus the inverse
wavelength $\xi/\lambda_{cont}(n,m)$ where the wavelength is
inferred from Eq.~(~\ref{eq:lambda}) for the quantum numbers $n$
and $m$ associated with the mode index $p$. Again we set (to some
extend arbitrarily) $\xi\equiv 30 a$.
As can be seen, all the data sets obtained for various sizes
collapse and confirm within numerical accuracy the choice of the
scaling variables. Interestingly, {\em two} separate  scaling
functions appear for transverse and longitudinal modes and, for
clarity, we have plotted both in two different graphs. The
crossover occurs at about $\lambda_{cont} \approx \xi$ for the
transverse modes in agreement with the observations in
Fig.~\ref{figEWfs}(b). In contrast, about twice as large
wavelengths are required for longitudinal modes to obtain a
satisfactory match with continuum theory as shown in
Fig.~\ref{figyscal}(b). Both scaling curves are similar
nevertheless and could indeed be brought to collapse by choosing a
larger $\xi$ for the longitudinal waves.
That $\xi$ depends somewhat on the type of mode is not surprising.
However, it would be of course more appealing if one would have a simple
physical argument explaining the {\em slower} crossover for longitudinal waves.

\subsection{Eigenvectors: Noise and Correlations.}

Snapshots similar to those presented in fig.~\ref{figEVagg} could
be presented for periodic bulk systems, however, as they turn out
to be intricate linear superpositions of plane waves and as the
existence of the continuum limit for the largest of our systems is
now sufficiently demonstrated we focus in the reminder of this
paragraph on the  {\em departure} from the continuum prediction.
%We have done this for configurations with $L=104$, $N=10,000$ and $\rho=0.925$
%as these systems are relatively large with their longest wavelengths well in the continuum
%branch of the crossover scaling function and since the ensemble is quite large ($M=20$)
%to obtain good statistics.

%\fbox{Fig. Projections of modes on elastic modes}

\begin{figure}
\centerline{
\epsfig{file=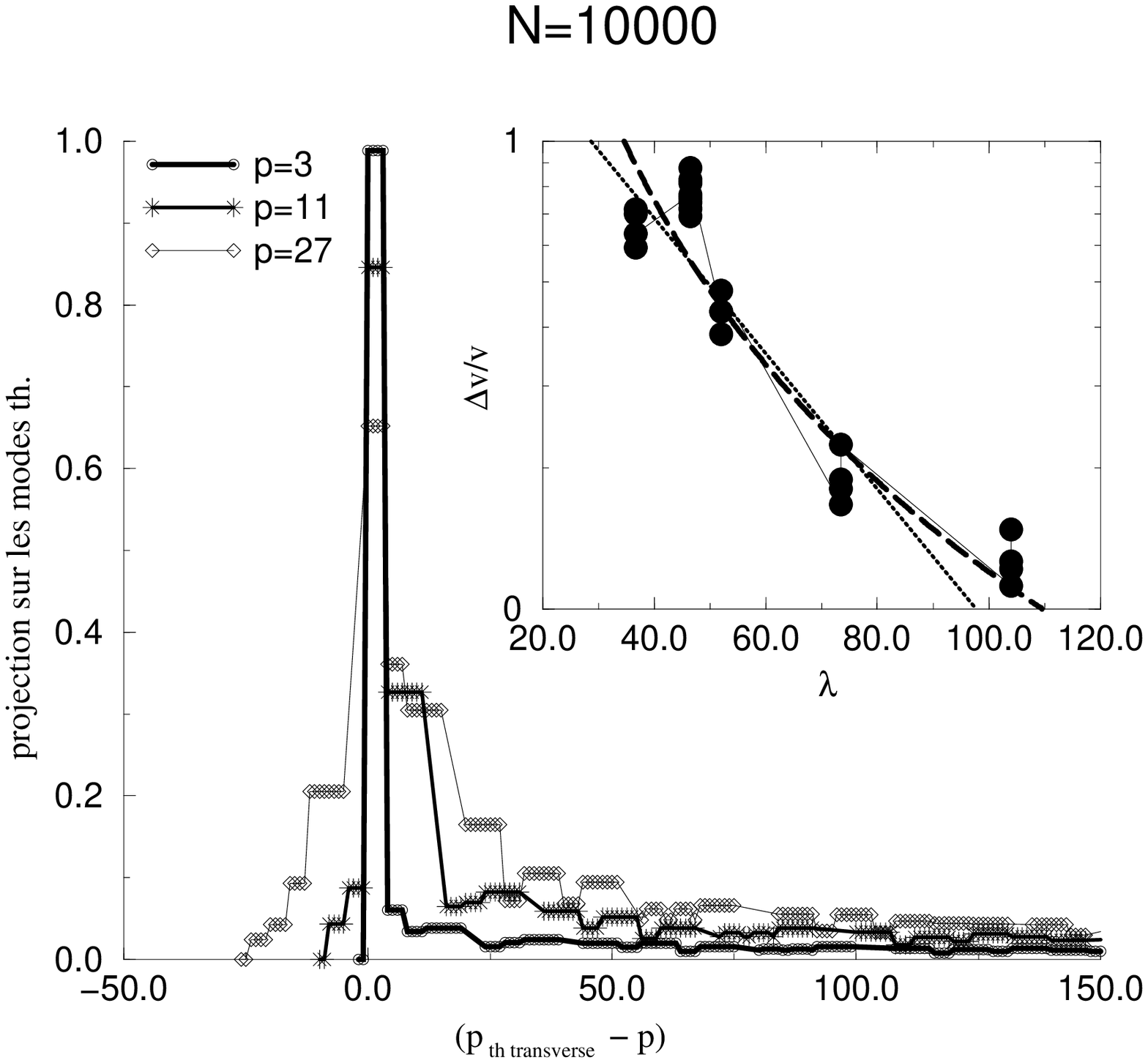,width=120mm,height=100mm}
}
\vspace*{0.5cm}
\caption{
Construction of noisy eigenvector field for periodic box configurations
with $N=10,000,\rho=0.925$.
Main figure:
projection amplitude of empirical eigenvectors $p=3,11$ and $27$ on the theoretical
plane waves which are indexed with $q$ with increasing frequency.
Only the transverse modes are included for clarity.
Insert: relative amplitude of noise as a function of wavelength $\lambda(p)$.
The dotted line is a fit with $\exp(-\lambda_{cont}/30a)$ in agreement with the
estimation $\xi\approx 30a$ for the characteristic wave length. 
The long-dashed line is a fit with $\lambda^{-2}$ in agreement with a scattering process.
\label{figEVproj}}
\end{figure}

In order to characterize the departure of the numerical
eigenvector displacement  fields $\vvec(p)$ from continuum theory
we project them onto plane waves $\vvec_{cont}(q)$, i.e. we
compute their Fourier amplitudes $A_p(q) = \langle \vvec(p) |
\vvec_{cont}(q)\rangle$. This is shown in Fig.~\ref{figEVproj} for
the eigenvectors $p=3$, $11$ and $27$. The average $\langle ... \rangle$
is taken over the ensemble. The amplitudes are plotted versus
$q-p$, i.e. we have shifted the abscissa axis  horizontally in
such a way as to emphasize the contribution of the $p$-th elastic
mode to the computed eigenvector with the same number. Indeed, the
main contribution is seen to be due to the plane wave
(propagative) mode with the same mode number.
The three particular eigenvectors considered in
Fig.~\ref{figEVproj}, belong to  sets of four-fold degenerate
eigenstates. Hence, if noise could be discarded the projections
onto plane waves  would be of width four, corresponding to an
average over all possible random phases. Accordingly, the
projection of eigenvectors belonging to an eight-fold degenerated
set would have a  width eight (not shown).
As anticipated by our discussion of the eigenvalues the overlap
between numerical and theoretical eigenmodes deteriorates with
increasing mode index $p$. This is seen from the decreasing peak
height and the increasing width of the function $A_p(q-p)$ with
increasing $p$. The enlargement of the peak to neighboring
frequencies suggests a scattering process in agreement
with~\cite{diffusion,sheng}. Note the asymmetric character of
the projections amplitudes,  which must  vanish for $q < 3$.
Interesting,  even  for small $p$, the amplitudes do not
completely vanish for large $q-p$, but become more or less
constant. This indicates a Fourier transformed {\em localized}
noise term, in agreement with the quasi-localized modes described
in Ref.~\cite{localization}. As we shall elaborate now below, this
is due to vortices in analogy to those depicted in
Fig.~\ref{fignoisefield}.

%\fbox{Description of how noise field is constructed.}

The next step consists in the construction of the noise field by
substracting the contributions of the dominant peak from the
numerical eigenvectors $\vvec(p)$:
\begin{equation}
\delta \vvec(p) \equiv \vvec(p) - \sum_{q \in {\cal D}_p } A_p(q) \vvec_{cont}(q)
\end{equation}
where ${\cal D}_p$ is the set of 4 (or 8) plane waves $q$ that
contribute most to the Fourier decomposition of the mode $p$.
Obviously, $||\delta \vvec(p)|| = \sqrt{ \sum_{q \notin {\cal D}_p} A_p(q)^2}$
should increase with the wavelength.
We have computed the dimensionless ratio $||\delta
\vvec(p)||/||\vvec(p)||$ and plotted this quantity in the inset of
Fig.~\ref{figEVproj} versus the wavelength $\lambda_{cont}(p)$.
This curve is in qualitative agreement with a scattering process
of Rayleigh type~\cite{sheng}, where $||\delta
\vvec(p)||/||\vvec(p)||\propto \lambda^{-2}$. We have compared the
data with the exponential decay $\exp(-\lambda_{cont}/30a)$.
Interestingly, the characteristic wavelength defined here is
equal to $\xi$. Our data do not allow to discriminate between
both fits. The conclusion is thus that, whatever the origin of the
noise (scattering process or not), the noise is small compared to
the propagative theoretical mode when $\lambda \gg\xi$.

%\fbox{Discussion of snapshots}

Let us now study the structure of the noisy part of the
wavevector. Assuming a scattering process, it would be
particularly interesting to determine the dependency of a possible
mean free path into the wavelength $\lambda_{cont}$ of the
eigenmode. Two examples for eigenvector noise fields $\delta
\vvec(p)$ are presented in the figures ~\ref{fignoisefield}(c) and
(d) for the modes $p=3$ and $p=7$ respectively. They are compared
with the non-affine fields obtained for the same configuration in
an elongational and pure shear displacement field. The vortices
are again the most striking features.
The four fields given look indeed remarkably similar: The sizes
and positions of the vortices are obviously highly correlated. To
put this in quantitative terms we  consider  correlation functions
for the eigenvector noise fields designed in analogy to those
discussed for the non-affine fields.
% and discuss then the
%connection between both sets of fields.

%\fbox{Fig. Correlation functions for eigenvector noise field}

The correlation function $C_v(r) \equiv \left< \delta \vvec(r) |
\delta \vvec(0) \right>$ is presented in Fig.~\ref{figvcorr} for
two modes $p=3$ and $p=11$. As expected from the noise field
snapshots we find again the anti-correlations similar to those
presented for the non-affine fields in Fig.~\ref{figbcorr}(a). The
anti-correlation extends even to somewhat larger distances. We
realize that the mode dependence while visible is weak. In the
inset we have plotted the first node of the anti-correlation
$\zeta_1(p)$. Also included is the wavelength corresponding to the
mode number $p$. We find that $\zeta_1$ does not vary much with
$\lambda$, unlike the $\lambda^4$-dependence of the mean free path
in scattering processes. Thus the noise displays a characteristic
length $\zeta_1$ comparable to $\xi$ and independent on the mode
$p$. We stress that the resulting participation ratio of the noise
is weak, in agreement with {\it localization}~\cite{localization}.
However, the slow (non exponential) decay of the correlation
function is in favor of {\it delocalization} as
in~\cite{ruocco2,radjai}.

\section{Discussions. }\label{sec:Discussion}

%\fbox{What we did}

In summary, we have presented extensive simulations of mechanical
and low-frequency vibrational properties of quenched amorphous
disk-shaped aggregates and periodic bulk systems. Two dimensional
ensembles containing up to $40,000$ polydisperse Lennard-Jones
particles have been generated and analysed in terms of sample
size, density, sample symmetry and quench protocol. We have
focused on systems with densities close to the zero-pressure state
to have similar conditions for the two boundary symmetries
studied.
The eigenmodes of the structures are calculated by diagonalization of the dynamical matrix
and the eigenfrequencies are compared with the predictions from classical continuum theory
where we concentrate on the low frequency end.
The second key calculation we performed consists in macroscopic
deformations (pure elongation and pure shear) of a periodic box in
order to obtain the elastic moduli (Lam\'e coefficients) and the
microscopic displacement fields. These are in turn compared with
the noisy part of the corresponding eigenvector fields.

%\fbox{key results}

The central results of this study are as follows:

1. The application of continuum elasticity theory is subject to strong limitations
in amorphous solids, for system sizes below a length scale $\xi$ of typically $30$
interatomic distances. This length scale is revealed in a systematic finite-size
study of the eigenfrequencies and in the crossover scaling of modes $p$ for fixed
system size with the wavelength $\lambda(p)$.

2. The success of the scaling of the eigenfrequencies with the
wavelength demonstrates that $\xi$ is within numerical accuracy
independent of the mode. The crossover behaviors of transverse and
longitudinal modes are similar, although the latter is somewhat
slower, i.e. slightly larger systems are required to match the
same mode with continuum theory.

3. The macroscopic deformation experiments demonstrate that the
non-affine displacements of the atoms on local scale matter: a
finite amount of energy is stored in the non-affine field and
there is a large difference between the true Lam\'e coefficients
and those obtained from the dynamical matrix, assuming an affine
displacement field on all scales.

4. Below a length scale similar to $\xi$ both the non-affine
displacement field and the noisy part of the eigenvector fields
displays  vortex like structures.  These structures are
responsible for the striking anti-correlations in the vectorial
pair correlation functions of both types of fields. We have
identified in this paper the non-affine field as the central
microscopic feature that makes the continuum approach
inappropriate.

%\fbox{Inhomogeneities of elastic moduli}

\begin{figure}
\centerline{
\epsfig{file=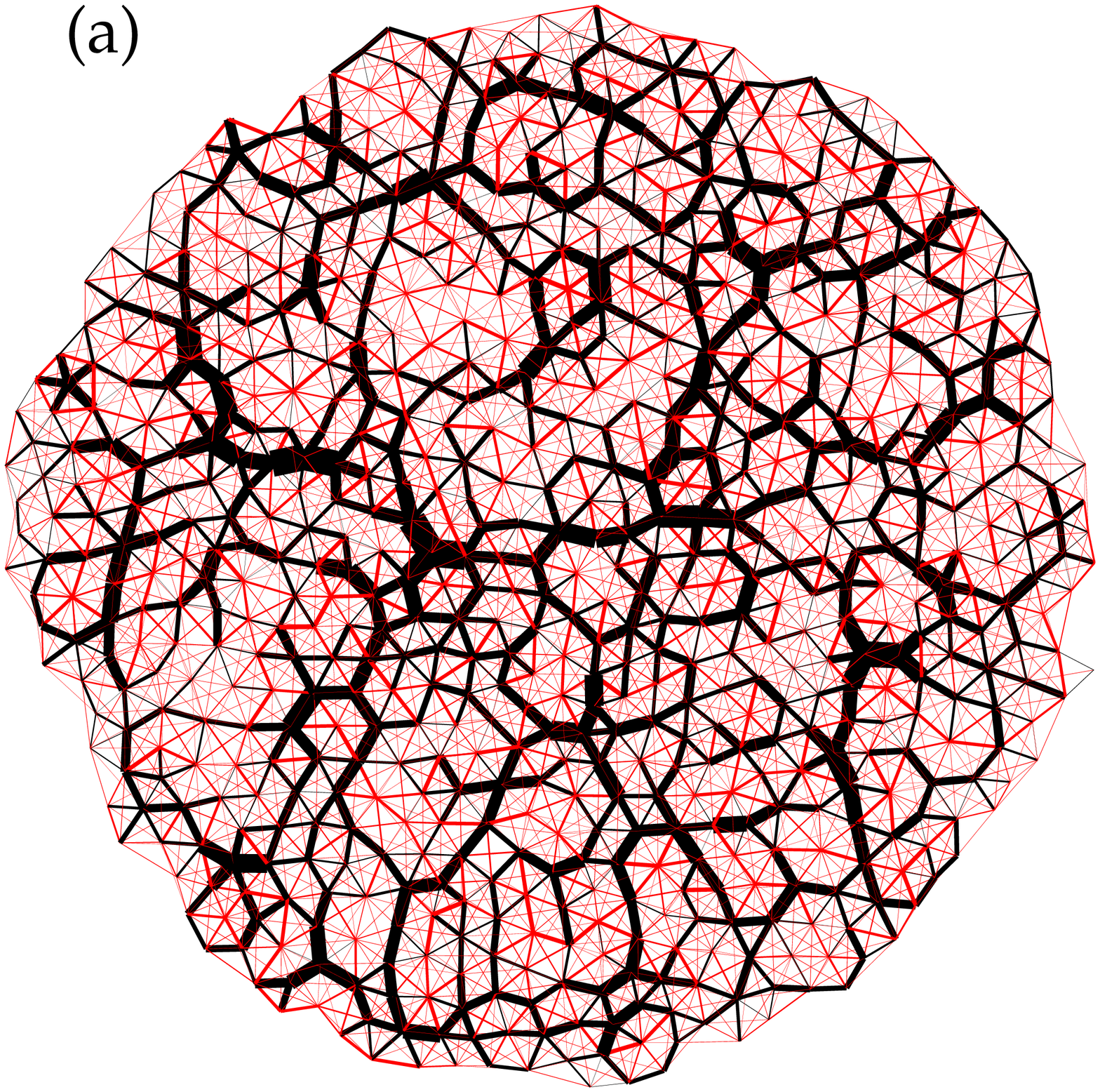,width=75mm,height=75mm}
\epsfig{file=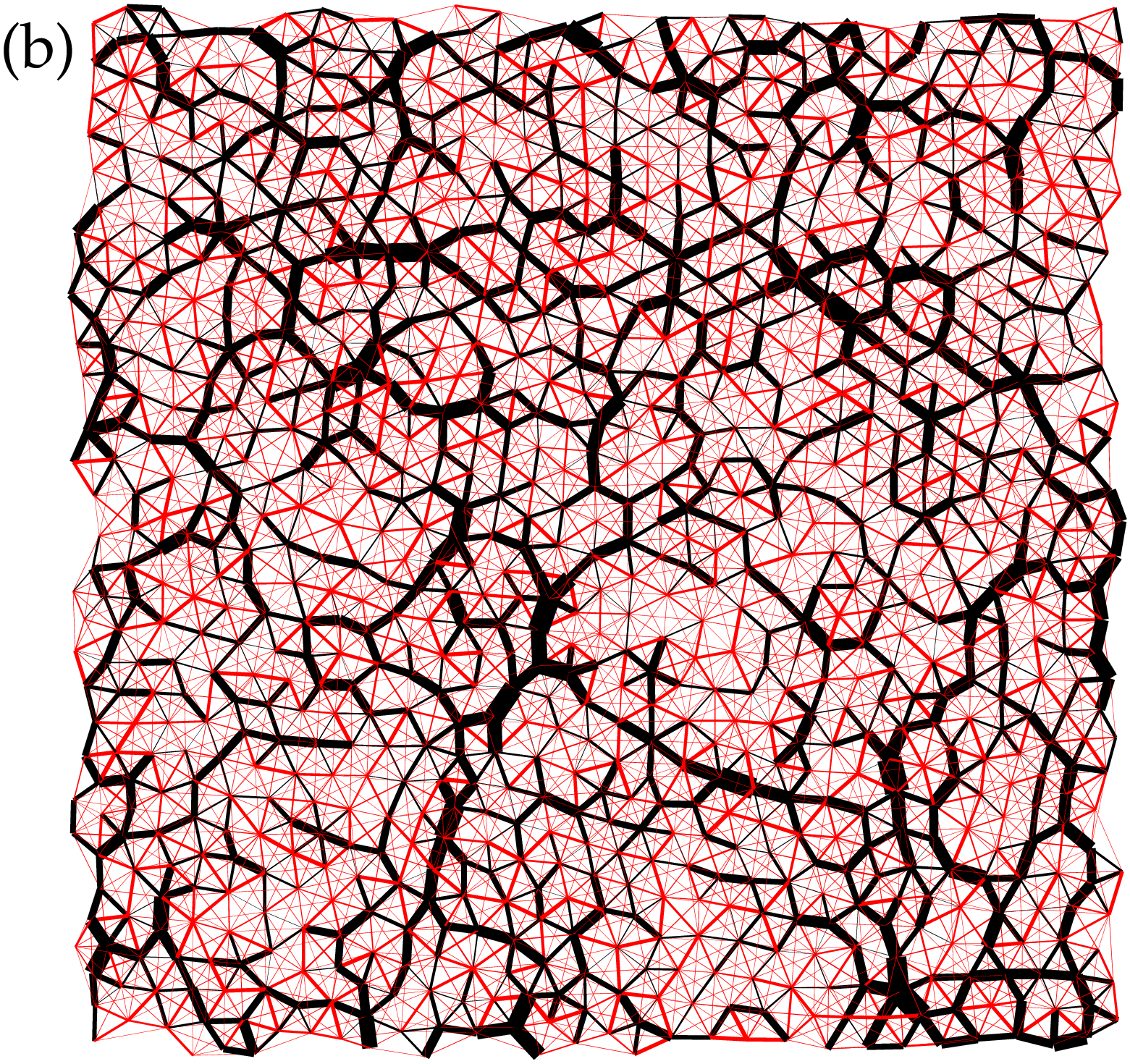,width=75mm,height=75mm}
}
\vspace*{1.5cm}
\caption{
Representation of the network of quenched stresses in two small
quenched Lennard-Jones particle systems in two dimensions:
{\bf (a)} a disk-shaped aggregate of diameter $2R\approx 32a$ containing $N=732$
particles (Protocol I) on the left and
{\bf (b)} a periodic bulk system with $L=32.9a$ and $N=1000$ (Protocol III)
on the right hand side.
The line scale is proportional to the tension transmitted along the
links between beads. The black lines indicate repulsive forces
(negative tensions), while the red links represent tensile forces between the verticies.
Both shown networks are very similar despite different symmetries and
quench protocols.
They are strongly inhomogeneous
and resemble the pattern seen in granular materials.
Zones of weak attractive links appear to be embedded within the strong skeleton
of repulsive forces.
\label{figFlink}}
\end{figure}

%
Inhomogeneities in local elastic coefficients, are an essential ingredient in
several recent calculations on disordered elastic systems \cite{Schirmacher,Ruocco}.
Indeed, the origin of the departure from elastic behavior seen in the two key
computer experiments is ultimately related to the local disorder.
This disorder is revealed by the structure of the force network frozen into the solid, as
shown in Fig.~\ref{figFlink}. Interestingly, weak finite-size effects are evidenced
in properties of the frozen local disorder like the pressure, the particle energy and
one of the Lam\'e coefficients and in the histograms of forces, coupling constants and
dynamical matrix trace.

Surprisingly at first sight, we have been unable to identify a sufficiently large
length scale solely from the frozen local disorder. There must be a mechanism
which amplifies these in such a way that that the deformation fields
as the ones displayed in figure \ref{fignoisefield} become non-affine
on scales comparable to $\xi=30a$.
Incidentally, this mechanism is not directly related to the mean
free path that can be computed in diffusion
processes~\cite{sheng} as can be shown by the absence of
any $\lambda$-dependence. One marked difference is that the mean
free path is obtained in one-dimensional
models~\cite{sheng,leibig}, however the characteristic length is
related in our study to vortices, implying a displacement in the
transverse direction. A one-dimensional model of the displacement
field in a two-dimensional medium as in Ref.~\cite{leibig} appears
thus to be unrealistic. Unfortunately, we are not able  at the
moment to propose a definite relation between the size of the
vortices and the local properties of the system. Preliminary
studies~\cite{inprogress} suggest a strong correlation between the
vortices and the occurrence under mechanical sollicitation of
nodes of stresses acting as bolts and forcing a displacement in
the transverse direction. This must be related to the local
anisotropy of forces as already been suggested in
Ref.~\cite{tanguyNoz} in an other context.

%\fbox{Boson peak}

Interestingly, sizes similar to $\xi$, or somewhat smaller, are
often invoked \cite{boson,glassyvibs1}, as typical of the
heterogeneities that give rise to anomalies in the vibrational
properties of disordered solids (glasses) in the Terahertz
frequency domain, the so-called boson peak.
In particular, Ref. \cite{boson} considers the
existence of rigid domains separated by softer interfacial zones,
not unlike those revealed by the non-affine displacement pattern of
Fig.~\ref{fignoisefield}(a). Our work offers a new vantage point on this
feature. At the wavelength corresponding to these Thz vibrations,
comparing the vibrational density of states to that of a
continuum, elastic model (the Debye model) is not necessarily meaningful.
%

%\fbox{Finite-Size Studies}

The present study documents the importance of systematic
finite-size characterizations for the computational investigations
of glassy systems well below the glass transition. It suggests
that numerical investigation of vibrational properties should
systematically make use of samples much larger than $\xi$, in
order to avoid finite-size effects and to be statistically
significant. Too small systems tend to be slightly more rigid and
have higher pressures and system energies.
%

%\fbox{Petites deformations versus ecoulements}

Finally, let us mention that such vortices have been studied in
disordered materials in the context of large deformations and
flow~\cite{radjai,langer}. We show here that the same mechanism
happens even in the elastic regime, for very small deformations in
a quasi-static motion. Vortex like deformation patterns have also
been identified, and associated with,  in simulations of granular
materials~\cite{granular}. Our studies clearly shows that very
simple models, involving only conservative forces, are sufficient
to reproduce such patterns. Disorder appears to be much more
relevant than the 'granular' aspect (e.g. frictional terms) for
this type of properties.

%\fbox{Natural extensions of this work}

The above conclusions  are obviously subject to the conditions
under which  our simulations have been performed. The most serious
limitation of this work is certainly that we have only reported
results on two dimensional samples. Of course, one expects
correlations to be reduced in higher dimensions and finite-size
effects should be less troublesome there.
However, this study initially originated from an attempt to compute the vibrational modes
of three dimensional clusters. Surprisingly, we have been unable to reach there the elastic
limit even for systems containing $10,000$ atoms and had to switch to the simpler two
dimensional case which is in terms of particle numbers less demanding
even though the length scale $\xi$
might ultimately turn out to be smaller in three dimensions.
Indeed, we believe that a systematic finite-size analysis of mechanical and vibrational
properties in three dimensional amorphous bodies is highly warranted and we are currently
pursuing simulations in this direction.

Other system parameters like the polydispersity index and
specifically the quench protocol should be systematically altered
to further the understanding of the origins of the demonstrated
correlations. We have not seen within the accuracy of our data any
systematic effect of the quench protocol on the the characteristic
length $\xi$.
 However, more attention should certainly be paid to the ageing processes
occurring in quenched samples as a function of system size.
 
\acknowledgments 
We thank U.~Buchenau, B.~Doliwa, E.~Duval, P.~Holdsworth, W.~Kob, and L.~Lewis 
for stimulating discussions. 
Generous grants of CPU time at the CDCSP Lyon are gratefully acknowledged. 
 
%%%%%%%%%%%%%%%%%%%%%%%%%%%%%%%%%%%%%%%%%%%%%%%%%%%%%%%%%%%%%%%%%%%%%%%%%%%%%%% 

\end{multicols} 

% Tabs
\newpage

% Protocol 1

\begin{table}[t]
\begin{tabular}{|l|c|c||c|c|c|c|c|c|c|}
$N$  & $M$ & $R_i/a$ & $R/a$ & $\rho$ & $e$ & $E$ &  $\lambda_a$ & $\mu_a$ 
& $\langle r_{ij}^2 c_{ij}\rangle$  \\ \hline
16	& 10&	 3&  	 2.3&	0.935&	0.58&	-2.08&	17.9&	20.6&	38.6 \\
32	& 10&	 4&	 3.4&	0.899&	0.57&	-2.31&	19.9&	20.1&	35.8 \\
64	& 11&	 5&	 4.7&	0.919&	0.55&	-2.47&	21.4&	23.2&	34.5 \\
128	& 20&	 7&	 6.7&	0.911&	0.41&	-2.58&	22.7&	22.6&	33.5 \\
256	& 10&	 8& 	 9.3&	0.948&	0.43&	-2.66&	24.8&	23.3&	32.8 \\ 
512	& 10&	11&     13.1&	0.954&	0.38&	-2.71&	25.1&	24.8&	32.5 \\
732	& 20&	15&	15.9&	0.920&	0.34&	-2.74&	24.2&	24.9&	32.4  \\	
1024	& 10&	19&	18.9&	0.910&	0.29&	-2.75&	24.3&	24.0&	32.3	\\
2048	& 10&	27&	26.7&	0.915&	0.19&	-2.78&	24.6&	24.5&	32.1	\\
4096	& 10&	38&	37.7&	0.918&  0.15&	-2.80&	24.8&	25.1&	32.0   \\
8192	&  5&	54&	53.5&	0.912&	0.25& 	-2.81&	24.8&	24.9&	31.9  \\
10000	&  6&	60&	59.1&	0.912&	0.21& 	-2.81&	24.8&	24.9&	31.9	\\	
16384	&  4&	85&	76.9&	0.882&	0.16&	-2.80&	23.9&	24.1&	32.0  \\
32768	&  1&	120&   109.9&	0.864&	0.11&	-2.80&  23.5&   23.1&	32.0  \\
\hline
\end{tabular}
\vspace*{0.5cm}
\caption{Some properties characterizing disk-shaped clusters
generated by a slow aggregation following protocol I
with MD steps at $T=0.1,0.05,0.01,0.005$ and $0.001$. 
We have indicated the particle size
$N$, the number of configurations in the ensemble $M$, 
the radius $R_i$ of the initial sphere, the mean radius $R$
of the final globule, the mean density $\rho$, the excentricity $e$
obtained from the inertia tensor of the cluster,
the interaction energy per particle $E$,
the Lam\'e coefficients $\lambda_a$ and $\mu_a$ obtained using
eqns.~\ref{eq:lambdamuaffine}
and the mean spring constant $\langle r_{ij}^2 c_{ij} \rangle$.
The mean tension $\langle r_{ij} t_{ij} \rangle$ and, hence,
the pressure are too tiny to be measured accurately. 
\label{tab:protI}}
\end{table}

% Protocol 2
\begin{table}[t]
\begin{tabular}{|l|c||c|c|c|c|c|c|c|}
$R_i/a$ & $M$ & $\langle N \rangle$ & $\rho$ & $e$ & $E$ &  $\lambda_a$ & $\mu_a$
& $\langle r_{ij}^2 c_{ij} \rangle$  \\ \hline
5   & 5 & 73  &	0.932&	0.35 &	-2.52&	21.9&23.5&	34.1	\\
7   & 5 & 142  &0.925&	0.27 &	-2.61&	22.9&23.5&	33.4	\\
11  & 5 & 350&	0.925&	0.20 &	-2.70&	23.7&24.5&	32.7	\\
19  & 5 & 1043&	0.920&	0.15 &	-2.76&	24.3&25.2&	32.3	\\
27  & 5 & 2112&	0.923&	0.12 &	-2.77&	24.8&24.8&	32.1	\\
38  & 5 & 4180&	0.921&	0.08 &	-2.81&	24.9&25.1&	32.0	\\
52  & 5 & 7848& 0.917&  0.08 &  -2.81&  24.9&24.9&	31.9	\\
60  & 8 & 10448& 0.919&  0.05 &  -2.81&  25.0&25.1&	31.9	\\
85  & 8 & 20968& 0.917&  0.06 &  -2.82&  25.1&25.0&	31.8	\\
120 & 4 &37030& 0.910&  0.04 &  -2.84&  25.9&25.9&	32.9   \\	
\hline
\end{tabular}
\vspace*{0.5cm}
\caption{Some properties characterizing disk-shaped clusters
generated with the fast quench protocol II which takes advantage
of already quenched periodic bulk systems (protocol III).
As in the first table we indicate 
the radius $R_i$ of the initial disk,
the number of configurations in the ensemble $M$,
the (mean) particle number, 
the mean density $\rho$, the excentricity $e$,
the interaction energy per particle $E$,
the Lam\'e coefficients $\lambda_a$ and $\mu_a$ 
and the mean spring constant $\langle r_{ij}^2 c_{ij} \rangle$.
The final cluster radius is not given as it is essentially identical to $R_i$.
Note that in the second protocol the particle number $N$ fluctuates (very weakly) around its 
mean value $\langle N \rangle$ while it is a constant operational parameter in the first protocol.
The excentricity is much smaller here than in the first more realistic protocol. 
The information contained in the 
last four columns is very similar to the one in the correponding columns
of table~\ref{tab:protI} and table~\ref{tab:protIII} despite the fact
that different quench protocols have been used.
\label{tab:protII}}
\end{table}

% Protocol 3
\begin{table}[t]
\begin{tabular}{|l|c|c|c||c|c|c|c|c|c|c|c|}
$N$  & $L/a$ & $\rho$ & $M$ & $E$ & $P$ & $\lambda$ & $\mu$ &  $\lambda_a$ & $\mu_a$
& $\langle r_{ij} t_{ij} \rangle$ & $\langle r_{ij}^2 c_{ij} \rangle$  \\ \hline
50   &	 7.4&	0.925&20&	-2.77&	 1.34&	50.3&  9.7 &	25.8&	24.1&	-0.39&	38.3	\\
100  &	10.4&	0.925&20& 	-2.81&	 0.51&	43.1&  11.4&	23.9&	23.7&	-0.15&	34.0	\\
200  &	14.7& 	0.925&20&	-2.83&	 0.45&	40.9&  10.3&	24.2&	24.8&	-0.09&	33.1	\\
300  &	18.1&	0.925&20&	-2.83&	 0.43&	41.2&  11.0&	25.3&	25.7&	-0.13&	33.8	\\
500  &	23.3&	0.925&20&	-2.84&	 0.19&	39.3&  11.6&	24.3&	24.9&	-0.06&	32.6	\\
1000 &	32.9&	0.925&20&	-2.84&	 0.24&  39.6&  11.6&    25.3&   25.4&   -0.07&   32.9   \\
2000 &	46.5&	0.925&40&	-2.84&	 0.25&	40.1&  11.8&	25.8&	26.0&	-0.08&	32.9	\\
5000 &	73.5& 	0.925&20&	-2.84&	 0.29&	39.6&  11.3&	26.2&	26.5&	-0.09&	33.1 \\  \hline
10000&	110 &	0.826&1 &	-2.78&	-0.63&	14.5&  7.5 &	18.4&	18.4&	0.22&	28.2	\\
10000&  109 &   0.841&1 &	-2.81&	-0.61&	20.4&  8.3 &	20.0&	19.5&	0.21&	28.3	\\
10000&	108 &	0.857&1 &	-2.81&	-0.67&	21.2&  7.8 &	20.1&	19.7&	0.23&	28.0	\\
10000&	107 &	0.873&1 &	-2.81&	-1.06&	23.1&  8.0 &	18.9&	17.9&	0.36&	25.8	\\
10000&	106 &	0.890&1 &	-2.82&	-1.13&	25.3&  8.9 &	18.0&	17.4&	0.38&	25.4		\\
10000&	105 &	0.907&1 &	-2.84&	-0.48&	31.0&  11.0&	22.4&	22.5&	0.16&	29.2	\\
10000&	104 &	0.925&20&	-2.84&	 0.25&	39.5&  11.7&	26.2&	26.4&	-0.19&	32.9	\\
10000&	103 &	0.943&1 &	-2.83&	 1.24&	47.4&  13.3&	31.3&	31.4&	-0.38&	37.8	\\
10000&	102 &	0.961&1 &	-2.78&	 2.71&	59.5&  14.9&	37.7&	39.3&	-0.77&	44.0	\\
10000&  101 &   0.980&1 &	-2.72&	 4.25&	69.7&  18.8&	44.7&	46.6&	-1.15&	50.1	\\
10000&	100 &	1.000&1 &	-2.62&	 5.96&	84.0&  19.0&	51.9&	54.8&	-1.55&	56.4	\\  \hline
20000& 147.1&	0.925&3 &	-2.84&	 0.35&	40.4&  12.0&	26.8&	27.0&	-0.11&	33.4	\\
40000&	208 &	0.925&2 &	-2.84&	 0.33&	-   &	-  &	26.7&	27.1&	-0.11&	33.3  \\
\hline
\end{tabular}
\vspace*{0.5cm}
\caption{Some properties characterizing periodic bulk systems generated following protocol III
with MD steps at $T=1.0,0.5,0.1,0.05,0.01,0.005$ and $0.001$. In addition to properties also
recorded in the previous tables we have included here 
the Lam\'e factors $\lambda$ and $\mu$ obtained from a macroscopic deformation
and the mean tension $\langle r_{ij} t_{ij} \rangle$.
Note that $\langle r_{ij}^2 c_{ij} \rangle \gg \langle r_{ij} t_{ij}\rangle$.
\label{tab:protIII}}
\end{table}

\newpage
%Fig.1

%Fig.2
\newpage

%Fig3
\newpage

%Fig.4
\newpage

%Fig.5
\newpage
\begin{figure}
%\centerline{
%\epsfig{file=fig5a.eps,width=80mm,height=80mm}
%\epsfig{file=fig5b.eps,width=80mm,height=80mm}
%}
%\centerline{
%\epsfig{file=fig5c.eps,width=80mm,height=80mm}
%\epsfig{file=fig5d.eps,width=80mm,height=80mm}
%}
\vspace*{0.5cm}
\caption{Comparison of non-affine displacement field $\delta \uvec(r)$
with eigenvector field $\delta \vvec_p(r)$ for periodic box of size $L=104a$
containing $N=10,000$ particles:
{\bf (a)} non-affine displacement field under elongation in $x$ direction,
{\bf (b)} same for plain shear using Lees-Edwards boundary conditions \protect\cite{AT},
{\bf (c)} eigenvector field for $p=3$ and
{\bf (d)} eigenvector field for $p=7$.
This confirms that the different noise fields are
{\em non-gaussian} and are highly correlated
in space and with respect to each other.
Detailed inspection shows eddies like in turbulent flow.
\label{fignoisefield}}
\end{figure}

%Fig.6
\newpage
\begin{figure}
\centerline{
\epsfig{file=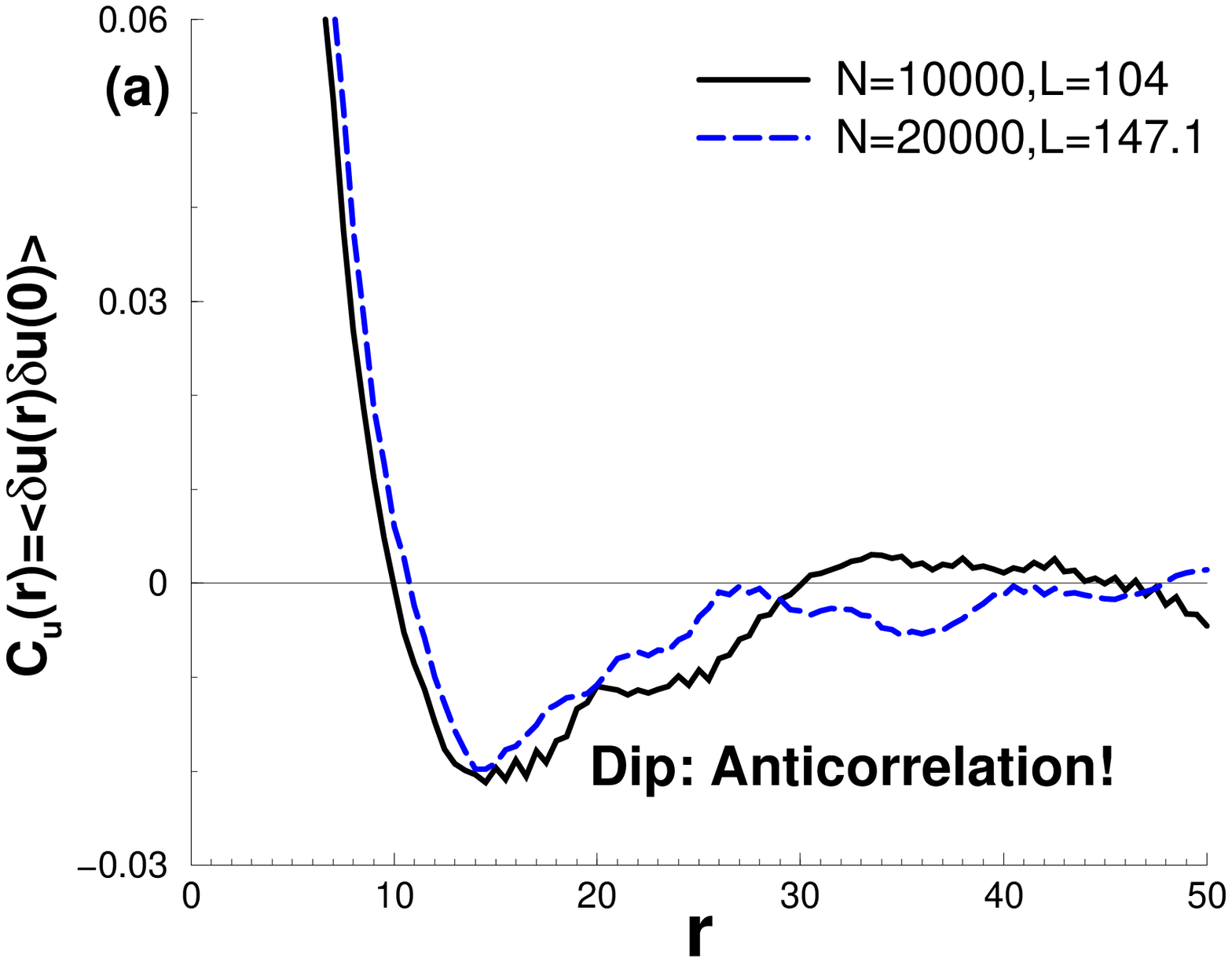,width=80mm,height=80mm}
\epsfig{file=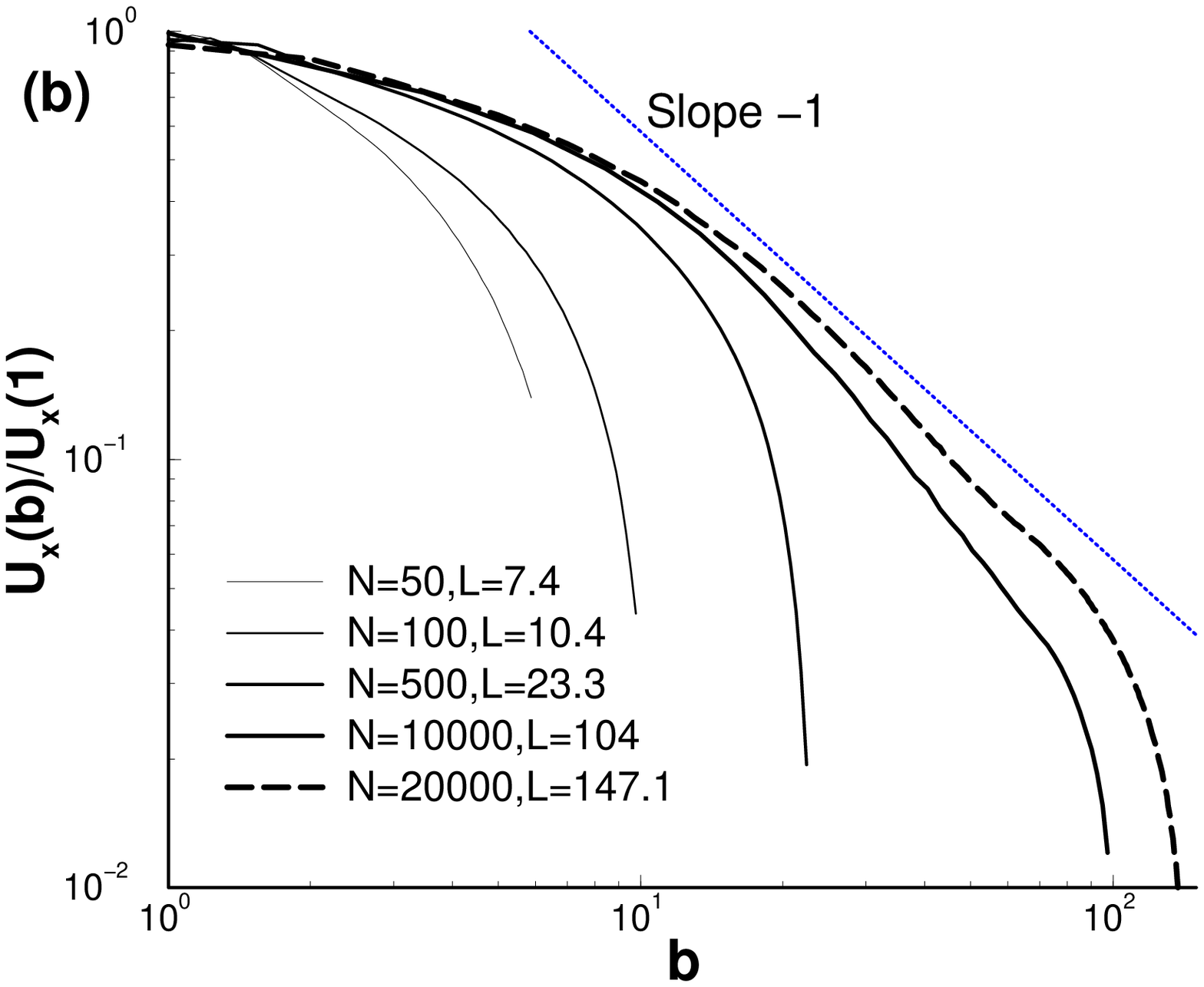,width=80mm,height=80mm}}
\vspace*{0.5cm}
\caption{
Characterization of non-affine displacement field $\delta u$-field obtained by simple elongation:
{\bf (a)} correlations function $C_u(r)=\langle \delta \uvec(r) \cdot \delta \uvec(0) \rangle$,
{\bf (b)} mean coarse-grained field $U_x(b)/U_x(b=1)$
{\em versus} the size $b$ of the coarse-graining.
Both functions become system size independent for large $L$.
The first figure on the left side shows clearly an anticorrelation in agreement with
the eddies seen in the snapshot fig.~\protect\ref{fignoisefield}(a).
\label{figbcorr}}
\end{figure}

%Fig.7
\newpage

%Fig.8
\newpage

%Fig.9
\newpage

%Fig.10
\newpage

%Fig.11
\newpage

%Fig.12
\newpage

%Fig.13
\newpage
\begin{figure}
\centerline{
\epsfig{file=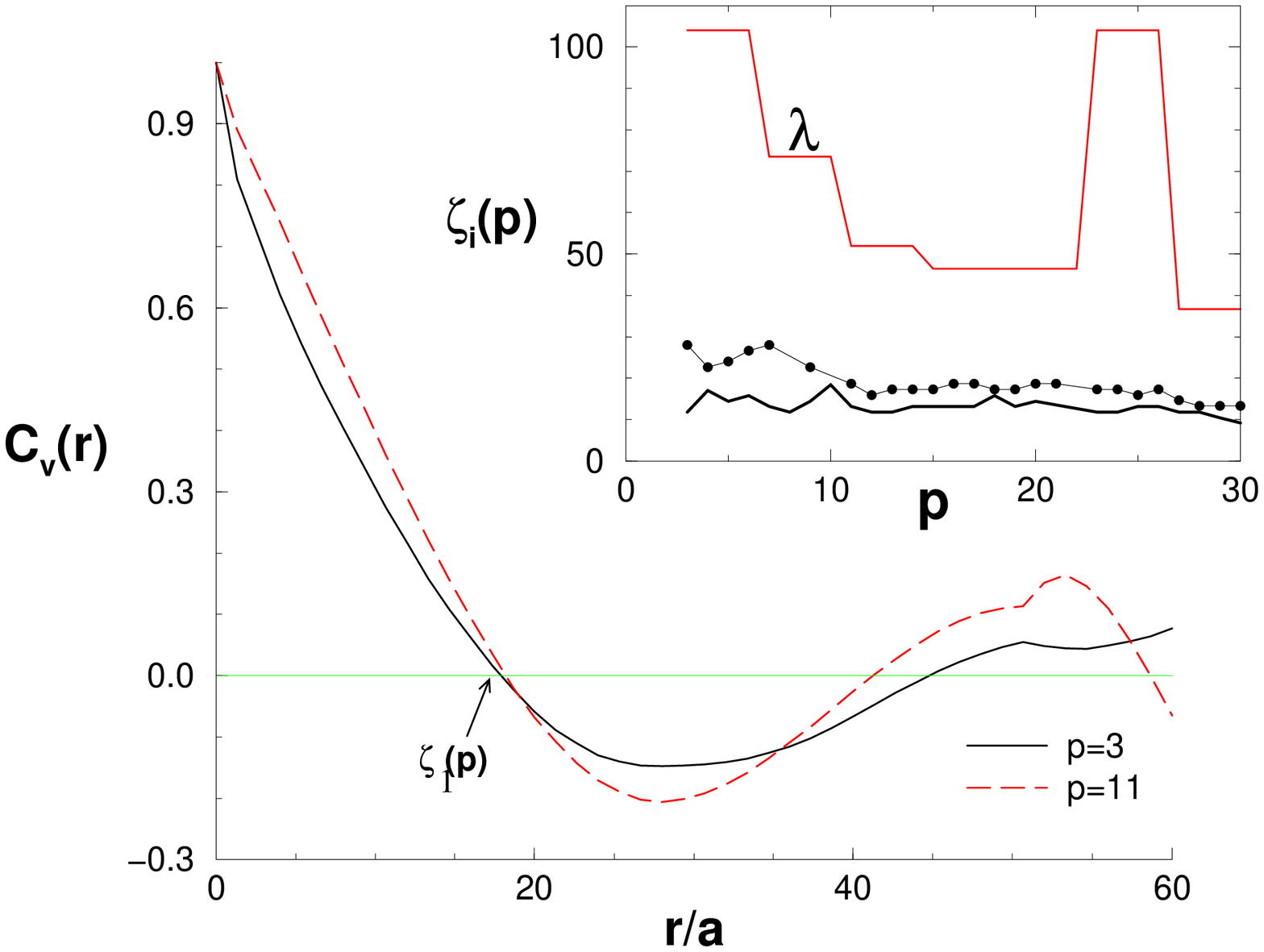,width=120mm,height=100mm}
}
\vspace*{0.5cm}
\caption{
Correlation function $\langle \delta \vvec_p(r) | \delta \vvec_p(0)  \rangle$
of the eigenvector noise fields for $p=3$ and $11$ {\em versus} distance $r$. The curves are similar
to the ones shown in fig.~\protect\ref{figbcorr}(b) and feature again prominent
anti-correlations.
Inset: $\zeta_1(p)$ characterizing the anti-correlation for N=10 000 (dotted line) and N=5 000 (bold line).
Also included is the wavelength from Eq.~\ref{eq:lambda} associated with each continuum mode
(top line).
\label{figvcorr}}

\end{figure}
 

\begin{thebibliography}{0}
\bibitem{Note} Note that both the wavelength and one of the
 Lam\'e coefficients are denoted by  $\lambda$.
To avoid ambiguity we write $\lambda(n,m)$, $\lambda(p)$ or $\lambda_{cont}$
for the wavelength where necessary.
\end{thebibliography}
\end{document}